\documentclass[showpacs,preprintnumbers,prd,nofootinbib,floats,amssymb,floatfix]{revtex4}
\usepackage{amsmath}
\usepackage{amsfonts}
\usepackage{graphicx}
\usepackage{hyperref}

\usepackage{amsmath}
\usepackage{amstext}
 
\begin{document}
\title{Faddeev-Jackiw quantization   of topological invariants: Euler and Pontryagin classes}
\author{Alberto Escalante}  \email{aescalan@sirio.ifuap.buap.mx}
\author{C. Medel-Portugal } \email{cmedel@ifuap.buap.mx}
 \affiliation{  Instituto de F{\'i}sica, Benem\'erita Universidad Aut\'onoma de Puebla,  \\
 Apartado Postal J-48 72570, Puebla Pue., M\'exico. }

\begin{abstract}
The symplectic analysis for the   four dimensional Pontryagin and Euler  invariants is performed within  the  Faddeev-Jackiw   context. The Faddeev-Jackiw constraints  and the generalized Faddeev-Jackiw brackets are reported;  we show that  in spite of the Pontryagin and Euler classes  give rise   the same equations of motion, its respective symplectic structures are different to each other. In addition,  a  quantum state that solves the Faddeev-Jackiw constraints is found,  and  we   show  that the  quantum states for these invariants  are different to each other. Finally,  we present some remarks and conclusions.
\end{abstract}
 \date{\today}
\pacs{98.80.-k,98.80.Cq}
\preprint{}
\maketitle
\section{INTRODUCTION}
Nowadays, the study of topological theories is an interesting topic  to perform. In fact,  the relevance for studying topological theories  has been motived in several contexts of theoretical physics because of they  provide an interesting  relation between  mathematics and physics,  just like that existing  between   geometry and General Relativity [GR]. From the classical point of view, topological theories are devoid of physical degrees of freedom,  background independent and diffeomorphisms covariant, because of these symmetries, the  topological theories are considered as good laboratories for testing ideas about the construction of  a background independent quantum   theory,  and these ideas could be applied for the construction of a    desired  quantum version of GR \cite{1}.  From a global  point of view, topology and quantum mechanics has an interesting overlap  just like that discovered  by  E. Witten  in \cite{2} and extended by M. Atiyah in \cite{3}, where concepts of  geometry, supersymmetry and quantum field theory  where unified, giving origin to the so-called  topological quantum field theory \cite{4}.  Moreover, we find in the literature that the  topological theories are also  important in the canonical approach of GR;  In fact,  when GR   is considered with  the addition of topological terms, namely  the Pontryagin, Euler and Nieh Yan invariants,  it is well-known that these topological invariants have
no effect on the equations of motion of gravity, however, they  give an important contribution in the symplectic structure of the theory \cite{5}. Within the  classical field theory  context,   either the    Euler  or Pontryagin classes   are   fundamental blocks for constructing the noncommutative form of topological gravity  \cite{6}.  Furthermore,   these topological invariants have been studied   in several   works due to they are expected to be related to physical observables, as for instance, in the case of anomalies  \cite{7, 8, 9, 10, 11, 12}.  It is important to comment, that the topological invariants cited above are not the only ones with an  interesting relation with physics. In fact, there are also the so-called  $BF$ theories \cite{13, 14}.  In general, a $BF$ theory is a topological theory,  it is  diffeomorphisms covariant,  and if  some   extra constraints  are imposed   on  the  $B$ field, then the topological structure of the theory  is  broken down  and theories  just as  complex GR \cite{15}, real GR \cite{ 16} or a  Yang-Mills theory arise in a  natural form \cite{17}. In addition, the   $BF$  formulations of gravity are interesting for the  community, because they have allowed  us  to understand    the spin foam formulation  developed  in the  Loop Quantum Gravity [LQG]  program  \cite{18}. In this respect,  both the   Euler  and   Pontryagin invariants can be written as a $BF$-like theory \cite{19}, and this important fact will be used along this paper. On the other hand, the  Euler  and  Pontryagin invariants are fundamental in the characterization of the topological structure of a manifold. In fact, they label topologically distinct four-geometries; the Pontryagin invariant gives the relation between the number of selfdual and anti-selfdual harmonic connections on the manifold. The Euler invariant, on the contrary,   gives a relation between the number of harmonic p-forms on the manifold \cite{20}. \\ 
From the Hamiltonian point of view, the Euler  and  Pontryagin invariants  treated  as field theories give rise   the same equations of motion, are devoid of physical degrees of freedom,  background independent,  diffeomorphisms covariant and there exist reducibility conditions between the constraints \cite{21}. Because of these symmetries, either   the Pontryagin or Euler invariants are good toy models for studying the classical and quantum structure of a  background independent theory.    \\
With these antecedents,   the purpose of this paper is to develop the symplectic analysis of the Euler  and  Pontryagin invariants. As far as we know the symplectic analysis of these invariants  has not been carryout. In this respect,  we have commented previously that these invariants has been analyzed only within the Dirac context in \cite{19};  however,  in these works the complete structure of the constraints and the  Dirac  brackets, useful  for the quantization of the theory,  were not constructed. In this manner, in order  to know a complete canonical description of the theories under study,  we   apply  the symplectic formulation of Faddeev-Jackiw [FJ] which is  a powerful alternative framework for studying singular systems  \cite{22}.  In fact, the FJ method is a symplectic approach where all relevant information of the theory can be obtained through an invertible symplectic matrix, which is constructed by means of the symplectic variables that are identified as the dynamical variables from  the Lagrangian of the theory. Since the theory is singular, there will be constraints and FJ scheme  has the advantage that all constraints are considered  at the same footing. In fact, in FJ method  it is not necessary to perform the classification of the constraints in primary, secondary, first class or second class, as it is done in Dirac's method. Furthermore, from the components of the symplectic tensor  it is possible to identify   the FJ generalized brackets;  one  goal of this  approach  is that,  at the end of the calculations,  the Dirac  and the  FJ generalized  brackets are equivalents. In this manner,   the cornerstone of this paper is to develop the symplectic analysis of the  Pontryagin and Euler invariants. In our results we will find that in spite of the Pontryagin and Euler classes  giving  rise   the same equations of motion, their  respective symplectic structures are different and this fact will be important in the quantization. In fact, once we have found the complete set of FJ constraints, we will find a quantum state that solves the constraints and we will see that the quantum states for these  invariants are different to each other. \\
The paper is organized as follows. In Section I,  the  symplectic formalism for the Pontryagin invariant is performed.  For our study, we will write the invariant as a $BF$-like theory, this fact  is necessary  because the FJ formalism is applicable to linear Lagrangians in the velocities.   We report the  complete set of FJ constraints and we will report the  reducibility conditions among these constraints. Then, a symplectic tensor is constructed and the FJ generalized brackets are identified. With  these results at hand,  we will find a quantum state that  solves the quantum FJ constraints. In Section  II   the symplectic analysis for the Euler invariant is  carried  out. For our analysis, we will use the same symplectic variables than those used in the Pontryagin invariant and we will show that despite  this fact  the symplectic structures are different to each other. Because of the symplectic structures are different, we will find a different quantum state that solves exactly the Euler's  constraints. Finally, we add some remarks and conclusions.

\section{Symplectic formalism for the Pontryagin invariant }
The four-dimensional Pontryagin invariant is described by the  action
\begin{equation}
S[A_{\mu}^{\ IJ}]=\frac{\Xi}{2}\int_{M}^{}\left[F^{IJ} \wedge F_{IJ} \right] \label{chern1},
\end{equation}
 here, $\Xi$ is a constant, $M$ is the   space-time manifold   and $F$ is  straight  field of the Lorentz connection $A_{\beta}^{\ IJ}$  given by  $F_{ \ \ \alpha \beta}^{IJ}=\partial_{\alpha}A_{\beta}^{\ IJ}-\partial_{\beta}A_{\alpha}^{\ IJ}+A_{\alpha \ K}^{\ I}A_{\beta}^{\ KJ}-A_{\alpha \ K}^{\ J}A_{\beta}^{\ KI}$, the capital letters  are internal $SO(3,1)$ Lorentz  indices and run from  $I,J,K=0,1,2,3$  that can be raised  and lowered by the internal metric $\eta^{IJ}=(-1,1,1,1)$, and $\alpha, \beta, \mu=0,1,2,3$ are space-time indices.\\
We  introduce   auxiliary fields, namely  $B^{IJ}$, corresponding to a  set of six two-forms; thus, the action (\ref{chern1})   takes a different fashion, a $BF$-like theory 
\begin{equation}
S[A_{\mu}^{\ IJ},B_{\alpha \beta}^{KL}]=\Xi \int_{M}^{}\left[F^{IJ}\wedge B_{IJ}-\frac{1}{2}B^{IJ} \wedge B_{IJ} \right] \label{chern2}.
\end{equation}
We can see that the action (\ref{chern1}) and (\ref{chern2}) are the same modulo equations of motion. On the other hand, with the introduction of the $B's$ variables, the action (\ref{chern2}) is now linear in the velocities, then the FJ formalism can be carried out \cite{22}. Furthermore, we will work with real variables without  involve either self-dual or anti-self-dual variables;  it is easy to observe that in the self-dual (anti-self-dual) scenario, the actions are  reduced  to the Pontryagin characteristic based on the self-dual (anti-self-dual) connection and this case is trivial.  \\
By performing the 3+1 decomposition and breaking down the Lorentz  covariance we obtain the following Lagrangian density
\begin{eqnarray}
L&=&\int_{M}^{}\Xi \eta^{abc}\Big[B^{0j}_{\ \ bc}\dot{A}_{a0j}+\frac{1}{2}B^{ij}_{\ \ bc}\dot{A}_{aij}+\frac{1}{2}A_{0ij}\left(\partial_a B^{ij}_{\ \ bc}+2B^{il}_{\ \ bc}A_{a \ l}^{\ j}+2B^{0i}_{\ \ bc}A_{a0}^{\ \ j} \right) \nonumber \\
&&+A_{00i}\left(\partial_a B^{0i}_{\ \ bc}+B^{ij}_{\ \ bc}A_{a0j}+B^{0j}_{\ \ bc}A_{a \ j}^{\ i} \right) \nonumber \\
&&+B^{0i}_{\ \ 0c}\left(\partial_a A_{b0i}-\partial_b A_{a0i}+A_{a0j}A_{b \ i}^{\ j}+A_{b0}^{\ \ j}A_{aij} \right) \nonumber \\
&& +\frac{1}{2}B^{ij}_{\ \ 0c}\left(\partial_a A_{bij}-\partial_b A_{aij}+A_{ai0}A_{b \ j}^{\ 0}+A_{ail}A_{b \ j}^{\ l}-A_{aj0}A_{b \ i}^{\ 0}-A_{ajl}A_{b \ i}^{\ l} \right) \nonumber \\
&& -\frac{1}{2}B^{0j}_{\ \ 0a}B_{0jbc}-\frac{1}{2}B^{0j}_{\ \ ab}B_{0j0c}-\frac{1}{4}B^{ij}_{\ \ 0a}B_{ijbc}-\frac{1}{4}B^{ij}_{\ \ ab}B_{ij0c}\Big]d^{3}x,\label{lagrangianochern}
\end{eqnarray}
here $a,b,c=1,2,3$, $\epsilon^{0abc}\equiv \eta^{abc}$ and $i, j, k=1,2,3$ are the internal  indices that can be raised or lowered with the Euclidean metric $\eta^{ij}=(1,1,1)$. By introducing the following variables
\begin{eqnarray}
&&A_{aij} \equiv -\epsilon_{ijk}A_{a}^{\ k}, \nonumber \\
&&A_{0ij} \equiv -\epsilon_{ijk}A_{0}^{\ k}, \nonumber \\
&&B_{abij} \equiv -\epsilon_{ijk}B_{ab}^{\ \ k}, \nonumber \\
&&B_{0aij} \equiv -\epsilon_{ijk}B_{0a}^{\ \ k}, \nonumber \\
&&A_{ai} \equiv \Upsilon_{ai}, \label{variables11}
\end{eqnarray}
the Lagrangian density takes the following form 
\begin{eqnarray}
\mathcal{L}&=&\Xi\eta^{abc}\left(B_{ab}^{\ \ 0i}\dot{A}_{c0i}+B_{abi}\dot{\Upsilon}_{c}^{\ i} \right)\nonumber \\
&&+A_{0}^{\ k}\left[\partial_c \left(\Xi\eta^{abc}B_{abk} \right)+\Xi\eta^{abc}\epsilon_{jkm}B_{ab}^{\ \ j}\Upsilon_{c}^{\ m}-\Xi\eta^{abc}\epsilon_{ijk}B_{ab}^{\ \ 0i}A_{c0}^{\ \ j} \right] \nonumber \\
&&+A_{00i}\left[\partial_c \left(\Xi\eta^{abc} B_{ab}^{\ \ 0i} \right)-\Xi\eta^{abc}\epsilon^{ij}_{\ \ k}A_{c0j}B_{ab}^{\ \ k}-\Xi\eta^{abc}\epsilon^{i}_{\ jk}B_{ab}^{\ \ 0j}\Upsilon_{c}^{\ k} \right]\nonumber \\
&&+\Xi\eta^{abc}B_{0a}^{\ \ 0i}\left[\partial_b A_{c0i}-\partial_c A_{b0i}+\epsilon_{i}^{\ jk}A_{b0j}\Upsilon_{ck}-\epsilon_{ijk}A_{c0}^{\ \ j}\Upsilon_{b}^{\ k} \right]\nonumber \\
&&+\Xi\eta^{abc}B_{0a}^{\ \ k}\left[\partial_b \Upsilon_{ck}-\partial_c \Upsilon_{bk}-\epsilon^{ij}_{\ \ k}A_{b0i}A_{c0j}+\epsilon_{kjm}\Upsilon_{b}^{\ j}\Upsilon_{c}^{\ m} \right] \nonumber \\
&&-\Xi\eta^{abc}\left[B_{0c}^{\ \ 0j}B_{ab0j}+B_{ab}^{\ \ k}B_{0ck} \right].\label{cherninicio}
\end{eqnarray}
In this manner, from (\ref{cherninicio})   the following symplectic form  of the  Lagrangian  is identified  \cite{22}
\begin{equation}
\mathcal{L}^{(0)}=\Xi\eta^{abc}B_{ab}^{\ \ 0i}\dot{A}_{c0i}+\Xi\eta^{abc}B_{abi}\dot{\Upsilon}_{c}^{\ i}-\mathcal{V}^{(0)}, \label{lagrangianosimplectico}
\end{equation}
where $\mathcal{V}^{(0)}$ corresponds  to the  symplectic potential expressed by 
\begin{eqnarray}
\mathcal{V}^{(0)}&=&-A_{0}^{\ i}\left[\partial_c(\Xi \eta^{abc}B_{abi} )+\Xi\eta^{abc}\epsilon^{j}_{\ ik}B_{abj}\Upsilon_{c}^{\ k}-\Xi\eta^{abc}\epsilon_{jki}B_{ab}^{\ \ 0j}A_{c0}^{\ \ k}\right]\nonumber \\
&&-A_{00i}\left[\partial_c(\Xi \eta^{abc}B_{ab}^{\ \ 0i} )-\Xi \eta^{abc}\epsilon^{i}_{\ jk}B_{ab}^{\ \ 0j}\Upsilon_{c}^{\ k}-\Xi \eta^{abc} \epsilon^{ijk}B_{abk}A_{c0j} \right]
 \nonumber \\
&&-\Xi \eta^{abc}B_{0a}^{\ \ 0i}\left[\partial_b A_{c0i}-\partial_c A_{b0i}+\epsilon_{i}^{\ jk}A_{b0j}\Upsilon_{ck}-\epsilon_{ijk}A_{c0}^{\ \ j}\Upsilon_{b}^{\ k} \right] \nonumber \\
&&-\Xi \eta^{abc}B_{0ai}\left[\partial_b \Upsilon_{c}^{\ i}-\partial_c \Upsilon_{b}^{\ i}+\epsilon^{i}_{\ jk}\Upsilon_{b}^{\ j}\Upsilon_{c}^{\ k}-\epsilon^{ijk}A_{b0j}A_{c0k} \right] \nonumber \\
&&+\Xi \eta^{abc}\left[B_{0a}^{\ \ 0i}B_{bc0i}+B_{bc}^{\ \ i}B_{0ai} \right].
 \end{eqnarray}
 From the symplectic Lagrangian (\ref{lagrangianosimplectico}) we identify the following symplectic variables
 \begin{equation}
\xi^{(0)}=(A_{a0i},B_{ab}^{\ \ 0i},\Upsilon_{a}^{\ i},B_{abi},A_{0}^{\ i},A_{00i},B_{a0}^{\ \ 0i},B_{0ai}),
\end{equation} 
and the following 1-forms
\begin{equation}
a^{(0)}=(\Xi\eta^{abc}B_{ab}^{\ \ 0i},0,\Xi\eta^{abc}B_{abi},0,0,0,0,0 ).
\end{equation} 
In this manner, the symplectic matrix defined as
$f_{ij}(x,y)=\frac{\delta a_j (y)}{\delta \xi^{i}(x)}-\frac{\delta a_{i}(x)}{\delta \xi^{j}(y)}$ \cite{22}, is given by

\begin{eqnarray}
f^{(0)}_{ij}=\left( \begin{array}{l l l l l l l l}
0 & \Xi \eta^{dec}\delta^{i}_{l} & 0 & 0 & 0 & 0 & 0 & 0\\ 
 \Xi \eta^{dec}\delta^{i}_{l} & 0 & 0 & 0 & 0 & 0 & 0 & 0\\
 0 & 0 & 0 & -\Xi\eta^{dec}\delta^{i}_{l} & 0 & 0 & 0 & 0\\
 0 & 0 & \Xi \eta^{dec}\delta^{i}_{l} & 0 & 0 & 0 & 0 & 0\\
 0 & 0 & 0 & 0 & 0 & 0 & 0 & 0\\
 0 & 0 & 0 & 0 & 0 & 0 & 0 & 0\\
 0 & 0 & 0 & 0 & 0 & 0 & 0 & 0\\
 0 & 0 & 0 & 0 & 0 & 0 & 0 & 0
\end{array} \right)\delta^{3}(x-y). \label{matrizf0}
\end{eqnarray}
We can observe that $f_{ij}^{(0)}$ is singular,  and therefore, there are  constraints. In order to identify the constraints, we calculate the zero-modes of $f_{ij}^{(0)}$  and they are  given by the following 4 vectors
\begin{eqnarray}
&&\nu^{(0)}_1=(0,0,0,0,V^{A_{0}^{\ i}},0,0,0),\\
&&\nu^{(0)}_2=(0,0,0,0,0,V^{A_{a0i}},0,0),\\
&&\nu^{(0)}_3=(0,0,0,0,0,0,V^{B_{a0}^{\ \ 0i}},0),\\
&&\nu^{(0)}_4=(0,0,0,0,0,0,0,V^{B_{a0i}}),
\end{eqnarray}
where $V^{A_{0}^{\ i}}$, $V^{A_{a0i}}$, $V^{B_{a0}^{\ \ 0i}}$ and $V^{B_{a0i}}$ are arbitrary functions. Hence, by using these modes we find the following FJ constraints
\begin{eqnarray}
\Omega^{(0)}_{i}&=&\int_{}^{}\mathbf{d}^{3}x\nu^{(0) \ i}_1\frac{\delta}{\delta \xi^{(0) \ i}}\int_{}^{}\mathbf{d}^{3}y \mathcal{V}^{(0)}(\xi)=\int_{}^{}\mathbf{d}^{3}x V^{A_{0}^{\ i}}\frac{\delta}{\delta A_{0}^{\ i}}\int_{}^{}\mathbf{d}^{3}y \mathcal{V}^{(0)}(\xi)\nonumber \\
&=&\partial_c(\Xi \eta^{abc}B_{abi} )+\Xi\eta^{abc}\epsilon^{j}_{\ ik}B_{abj}\Upsilon_{c}^{\ k}-\Xi\eta^{abc}\epsilon_{jki}B_{ab}^{\ \ 0j}A_{c0}^{\ \ k}, \label{restriccion1}\\
\Omega^{(0) \ 00i}&=&\int_{}^{}\mathbf{d}^{3}x\nu^{(0) \ i}_2\frac{\delta}{\delta \xi^{(0) \ i}}\int_{}^{}\mathbf{d}^{3}y \mathcal{V}^{(0)}(\xi)=\int_{}^{}\mathbf{d}^{3}xV^{A_{a0i}}\frac{\delta}{\delta A_{a0i}}\int_{}^{}\mathbf{d}^{3}y \mathcal{V}^{(0)}(\xi) \nonumber \\
&=&\partial_c(\Xi \eta^{abc}B_{ab}^{\ \ 0i} )-\Xi \eta^{abc}\epsilon^{i}_{\ jk}B_{ab}^{\ \ 0j}\Upsilon_{c}^{\ k}-\Xi \eta^{abc} \epsilon^{ijk}B_{abk}A_{c0j},\label{restriccion2}\\
\Omega^{(0) \ 0a}_{\ \ \ \ \ \ 0i}&=&\int_{}^{}\mathbf{d}^{3}x\nu^{(0) \ i}_3\frac{\delta}{\delta \xi^{(0) \ i}}\int_{}^{}\mathbf{d}^{3}y \mathcal{V}^{(0)}(\xi)=\int_{}^{}\mathbf{d}^{3}xV^{B_{a0}^{\ \ 0i}}\frac{\delta}{\delta B_{a0}^{\ \ 0i}}\int_{}^{}\mathbf{d}^{3}y \mathcal{V}^{(0)}(\xi) \nonumber \\
&=&\Xi \eta^{abc}\left[\partial_b A_{c0i}-\partial_c A_{b0i}+\epsilon_{i}^{\ jk}A_{b0j}\Upsilon_{ck}-\epsilon_{ijk}A_{c0}^{\ \ j}\Upsilon_{b}^{\ k}\right]-\Xi\eta^{abc}B_{ab0i},\label{restriccion3}\\
\Omega^{(0) \ a0i}&=&\int_{}^{}\mathbf{d}^{3}x\nu^{(0) \ i}_4\frac{\delta}{\delta \xi^{(0) \ i}}\int_{}^{}\mathbf{d}^{3}y \mathcal{V}^{(0)}(\xi)=\int_{}^{}\mathbf{d}^{3}x V^{B_{a0i}}\frac{\delta}{\delta B_{a0i}}\int_{}^{}\mathbf{d}^{3}y \mathcal{V}^{(0)}(\xi)\nonumber \\
&=&\Xi\eta^{abc}\left[\partial_b \Upsilon_{c}^{\ i}-\partial_c \Upsilon_{b}^{\ i}+\epsilon^{i}_{\ jk}\Upsilon_{b}^{\ j}\Upsilon_{c}^{\ k}-\epsilon^{ijk}A_{b0j}A_{c0k} \right]-\Xi \eta^{abc}B_{ab}^{\ \ i} \label{restriccion4}, 
\end{eqnarray}
from these constraints, we observe that there exist the following 6 reducibility conditions
\begin{eqnarray}
\partial_a \Omega^{(0) \ 0a}_{\ \ \ \ \ \ 0i}&=&\epsilon_{i}^{\ jk}\Upsilon_{ak}\Omega^{(0) \ 0a}_{\ \ \ \ \ \ 0j}+\epsilon_{i}^{\ jk}A_{a0k}\Omega^{(0) \ a0}{_{j}}+\frac{1}{2}\Omega^{(0) \ 00}{_{i}},  \nonumber \\
\partial_a \Omega^{(0) \ a0}{_{i}}&=&-\epsilon_{i}^{\ jk}A_{a0k}\Omega^{(0) \ 0a}_{\ \ \ \ \ \ 0j}+\epsilon_{i}^{\ jk}\Upsilon_{ak}\Omega^{(0) \ a0}{_{j}}+\frac{1}{2}\Omega^{(0)}_{i},
\end{eqnarray}
 and this fact will  be considered in the counting of physical degrees of freedom.  Now we shall observe if emerge   more constraints, for this aim, we calculate the following system \cite{22}
\begin{equation}
\bar{f}_{ij}\dot{\xi}^{(0) j}=Z_i(\xi),\label{terciarias}
\end{equation}
where
\begin{equation}
\bar{f}_{ij}=\left( \begin{array}{l}
f^{(0)}_{ij}\\
\frac{\delta \Omega_i^{(0)}}{\delta \xi^{(0) j}}
\end{array} \right)  \ \ \ \mathbf{y} \ \ \ Z_k=\left(\begin{array}{c}
\frac{\delta \mathcal{V^{(0)}}}{\delta \xi^{(0)}j}\\
0\\
0\\
0
\end{array} \right). \label{terciarias2}
\end{equation}
Thus, the symplectic matrix $\bar{f}_{ij}$ is given by
\begin{eqnarray}
\bar{f}_{ij}=\left(\begin{array}{c c c}
0 & \Xi \eta^{abc}\delta^{i}_{k} & 0 \\ 
 \Xi \eta^{abc}\delta^{i}_{k} & 0 & 0 \\
 0 & 0 & 0 \\
 0 & 0 & \Xi \eta^{abc}\delta^{i}_{k} \\
 0 & 0 & 0 \\
 0 & 0 & 0 \\
 0 & 0 & 0 \\
 0 & 0 & 0 \\
 -\Xi \eta^{abc}\epsilon_{j \ i}^{\ k}B_{ab}^{\ \ 0j} & -\Xi \eta^{abc}\epsilon^{\ j}_{k \ i}A_{c0j} & \Xi \eta^{abc}\epsilon^{j}_{\ ik}B_{abj}\\
 -\Xi\eta^{abc}\epsilon^{ikj}B_{abj} & \Xi\eta^{abc}\left(\delta_{k}^{i}\partial_c-\epsilon^{i}_{\ kj}\Upsilon_{c}^{\ j} \right) & -\Xi\eta^{abc}\epsilon^{i}_{\ jk}B_{ab}^{\ \ 0j}\\
 2\Xi\eta^{abc}\left(\delta_{i}^{k}\partial_c-\epsilon_{i}^{\ jk}\Upsilon_{bk} \right) & -\Xi\eta^{abc}\delta^{i}_{k} & 2\Xi\eta^{abc}\epsilon_{i \ k}^{\ j}A_{b0j}\\
-2\Xi\eta^{abc}\epsilon^{ijk}A_{b0j} & 0 & 2\Xi\eta^{abc}\left(\delta^{i}_{k}\partial_b+\epsilon^{i}_{\ jk}\Upsilon_{b}^{\ j} \right)
\end{array} \right.\nonumber 
\end{eqnarray}
\begin{eqnarray}
\left.\begin{array}{c c c c c}
0 & 0 & 0 & 0 & 0\\
0 & 0 & 0 & 0 & 0\\
-\Xi\eta^{abc}\delta^{i}_{k} & 0 & 0 & 0 & 0\\
0 & 0 & 0 & 0 & 0\\
\Xi\eta^{abc}\left(\delta_{k}^{i}\partial_c+\epsilon^{k}_{\ ij}\Upsilon_{c}^{\ j} \right) & 0 & 0 & 0 & 0\\
-\Xi\eta^{abc}\epsilon^{ijk}A_{c0j} & 0 & 0 & 0 & 0\\
0 & 0 & 0 & 0 & 0\\
-\Xi\eta^{abc}\delta^{i}_{k} & 0 & 0 & 0 & 0\\
\end{array}\right)\delta^3(x-y).
\end{eqnarray}
The matrix $\bar{f}_{ij}$ is not a square matrix as expected, however it has null vectors. The null vectors are given by
\begin{eqnarray}
&&\vec{\mathbb{V}}_1=(-\epsilon_{ijl}A_{c0}^{\ \ j}V^{l}\ ,\epsilon_{jil}B_{ab}^{\ \ 0j}V^{l}\ ,\partial_c V_i+\epsilon_{ijl}\Upsilon_{c}^{\ l}V^{j}\ ,-\epsilon^{j}_{\ li}B_{abj}V^{l},0,0,0,0,V^{i},0,0,0 ),\nonumber  \\
&&\vec{\mathbb{V}}_2= (\partial_c V_i-\epsilon^{l}_{\ ij}\Upsilon^{j}_{\ c}V_l \ ,\epsilon^{l \ j}_{i}B_{abj}V_l \ ,-\epsilon_{i}^{\ lj}A_{c0j}V_{l} \ ,\epsilon_{lji}B_{ab}^{\ \ 0j}V^{l},0,0,0,0,0,V^{i},0,0), \nonumber \\
&&\vec{\mathbb{V}}_3=(V_i\ ,-2(\partial_b V_i-\epsilon_{l}^{\ ji}\Upsilon_{bj}V^{l}) \ ,0,-2\epsilon_{l \ i}^{\ j}A_{b0j}V^{l},0,0,0,0,0,0,V^{i},0),\nonumber \\
&&\vec{\mathbb{V}}_4=(0,2\epsilon^{lj}_{\ \ i}A_{b0j}V_l \ ,-V_i\ ,-2(\partial_b V_i+\epsilon_{lji}\Upsilon^{j}_{\ b}V^{l}),0,0,0,0,0,0,0,V^{i} ),
\end{eqnarray}
where $V^{l}$'s  are arbitrary functions.
On the other hand, $Z_k(\xi)$ is given by
\begin{eqnarray}
Z_i(\xi)&=&\left(\begin{array}{c}
\frac{\delta \mathcal{V^{(0)}}}{\delta \xi^{(0)}j}\\
0\\
0\\
0
\end{array} \right)\\
&=&\left(\begin{array}{c}
\Xi\eta^{abc}\epsilon_{j \ i}^{\ k}A_{0}^{\ i}B_{ab}^{\ \ 0j}+\Xi\eta^{abc}\epsilon^{ikj}A_{00i}B_{abj}+2\Xi\eta^{abc}\partial_b B_{0a}^{\ \ 0k}\\
-2\Xi\eta^{abc}\epsilon_{i}^{\ kj}B_{0a}^{\ \ 0i}\Upsilon_{cj}+2\Xi\eta^{abc}\epsilon^{ijk}B_{0ai}A_{b0j}\\
\\
\Xi\eta^{abc}\epsilon_{ikj}A_{0}^{\ j}A_{c0}^{\ \ k}+\Xi\eta^{abc}\partial_cA_{00i}+\Xi\eta^{abc}\epsilon^{j}_{\ ik}A_{00j}\Upsilon_{c}^{\ k}\\
\\
-\Xi\eta^{abc}\epsilon_{\ ik}^{j}A_{0}^{\ i}B_{abj}+\Xi\eta^{abc}\epsilon^{i}_{\ jk}A_{00i}B_{ab}^{\ \ 0j}-2\Xi\eta^{abc}\epsilon_{i}^{\ jk}A_{b0j}B_{0a}^{\ \ 0i}\\
-2\Xi\eta^{abc}\partial_bB_{0ak}-2\epsilon^{i}_{\ jk}\Upsilon_{b}^{\ j}B_{0ai}\\
\\
\Xi\eta^{abc}\partial_cA_{0}^{\ i}-\Xi\eta^{abc}\epsilon^{i}_{\ jk}A_{0}^{\ j}\Upsilon_{c}^{\ k}+\Xi\eta^{abc}\epsilon^{jki}A_{00j}A_{c0k}\\
\\
\Omega^{(0)}_{i}\\
\\
\Omega^{(0) \ 00i}\\
\\
\Omega^{(0) \ 0a}_{\ \ \ \ \ \ 0i}\\
\\
\Omega^{(0) \ 0ai}\\
\\
0\\
\\
0\\
\\
0\\
\\
0
\end{array} \right).
\end{eqnarray}
In this manner, the contraction of the null vectors with $Z_k$, namely, $\vec{\mathbb{V}}_{i}^{\mu}Z_{\mu}(\xi)=0$, gives identities because the result  is a linear combination of constraints. Hence, there are no more FJ constraints.\\
Furthermore, we will add the constraints given in (\ref{restriccion1}-\ref{restriccion4}) to the symplectic Lagrangian using the following Lagrange multipliers, namely, 
$A_{0}^{\ i}=\dot{T}^{i},A_{00i}=\dot{\Lambda}_i,B_{0a}^{\ \ 0i}=\frac{\dot{\varsigma}_{a}^{\ i}}{2},B_{0ai}=\frac{\dot{\chi}_{ai}}{2} $, thus the symplectic Lagrangian reads
\begin{eqnarray}
\mathcal{L}^{(1)}&=&\Xi\eta^{abc}B_{ab}^{\ \ 0i}\dot{A}_{c0i}+\Xi\eta^{abc}B_{abi}\dot{\Upsilon}_{c}^{\ i}-\dot{T}^{i}\Omega^{(0)}_{i}-\dot{\Lambda}_i\Omega^{(0) \ 00i}-\frac{\dot{\varsigma}_{a}^{\ i}}{2}\Omega^{(0) \ 0a}_{\ \ \ \ \ \ 0i}\nonumber \label{lagrangianosimplectico2} \\
&&-\frac{\dot{\chi}_{ai}}{2}\Omega^{(0) \ 0ai}-\mathcal{V}^{(1)}\label{symplecticlagrangian1},
\end{eqnarray}
where
$\mathcal{V}^{(1)}=\mathcal{V}^{(0)}|_{\Omega^{(0)}_{i},\Omega^{(0) \ 00i},\Omega^{(0) \ 0a}_{\ \ \ \ \ \ 0i},\Omega^{(0) \ 0ai}=0}=0$. This result  is expected because of  the general covariance of the theory such as it is  present in GR.\\
From the symplectic Lagrangian (\ref{symplecticlagrangian1}) we identify the following symplectic variables
\begin{equation}
\xi^{(0)}=(A_{a0i},B_{ab}^{\ \ 0i},\Upsilon_{a}^{\ i},B_{abi},A_{0}^{\ i},A_{00i},B_{a0}^{\ \ 0i},B_{0ai},T^{i},\Lambda_i,\varsigma_{a}^{\ i},\chi_{ai} ),
\end{equation}
and the 1-forms
\begin{equation}
a^{(0)}=\left(\Xi\eta^{abc}B_{ab}^{\ \ 0i},0,\Xi\eta^{abc}B_{abi},0,-\Omega^{(0)}_{\ \ \ i},-\Omega^{(0) \ 00i},-\frac{\Omega^{(0) \ 0a}_{\ \ \ \ \ \ 0i}}{2},-\frac{\Omega^{(0) \ 0ai}}{2} \right).
\end{equation}
Hence, the symplectic matrix has the following form
\begin{eqnarray}
f^{(1)}_{ij}=\left(\begin{array}{c c c c}
0 & \Xi \eta^{abc}\delta^{i}_{k} & 0 & 0\\ 
 \Xi \eta^{abc}\delta^{i}_{k} & 0 & 0 & 0\\
 0 & 0 & 0 & -\Xi\eta^{abc}\delta^{i}_{k}\\
 0 & 0 & \Xi \eta^{abc}\delta^{i}_{k} & 0\\
 -\Xi \eta^{abc}\epsilon_{j \ i}^{\ k}B_{ab}^{\ \ 0j} & -\Xi \eta^{abc}\epsilon^{\ j}_{k \ i}A_{c0j} & \Xi \eta^{abc}\epsilon^{j}_{\ ik}B_{abj} & \Xi\eta^{abc}D^{i}
 _{kc}\\
 -\Xi\eta^{abc}\epsilon^{ikj}B_{abj} & \Xi\eta^{abc}d^{i}_{kc} & -\Xi\eta^{abc}\epsilon^{i}_{\ jk}B_{ab}^{\ \ 0j} & -\Xi\eta^{abc}\epsilon^{ijk}A_{c0j}\\
 \Xi\eta^{abc}d^{i}_{kc} & -\frac{\Xi}{2}\eta^{abc}\delta^{i}_{k} & \Xi\eta^{abc}\epsilon_{i \ k}^{\ j}A_{b0j} & 0\\
-\Xi\eta^{abc}\epsilon^{ijk}A_{b0j} & 0 & \Xi\eta^{abc}d^{i}_{kb} & -\Xi\eta^{abc}\delta^{i}_{k}
\end{array} \right. \nonumber
\end{eqnarray}
\begin{eqnarray}
\left. \begin{array}{c c c c}
\Xi\eta^{abc}\epsilon_{j \ i}^{\ k}B_{ab}^{\ \ 0j} & \Xi\eta^{abc}\epsilon^{ikj}B_{abj} & -\Xi\eta^{abc}d^{i}_{kc} & \Xi\eta^{abc}\epsilon^{ijk}A_{b0j}\\
\Xi\eta^{abc}\epsilon_{k \ i}^{\ j}A_{c0j} & -\Xi\eta^{abc}d^{i}_{kc} & \frac{\Xi}{2}\eta^{abc}\delta^{i}_{k} & 0\\
-\Xi\eta^{abc}\epsilon^{j}_{\ ik}B_{abj} & \Xi\eta^{abc}\epsilon^{i}_{\ jk}B_{ab}^{\ \ 0j} & -\Xi\eta^{abc}\epsilon_{i}^{\ jk}A_{b0j} & -\Xi\eta^{abc}d^{i}_{kb}\\
-\Xi\eta^{abc}D^{i}_{kc} & \Xi\eta^{abc}\epsilon^{ijk}A_{c0j} & 0 & \frac{\Xi}{2}\eta^{abc}\delta^{i}_{k}\\
0 & 0 & 0 & 0\\
0 & 0 & 0 & 0\\
0 & 0 & 0 & 0\\
0 & 0 & 0 & 0\\
\end{array}\right)\delta^3(x-y), \label{matrizsimplectica2}
\end{eqnarray}
where the notation
$D^{i}_{kc}=\delta^{i}_{k}\partial_c+\epsilon_{k}^{\ ij}\Upsilon_{cj}$ and $d^{i}_{kc}=\delta^{i}_{k}\partial_c-\epsilon_{k}^{\ ij}\Upsilon_{cj}$ was  introduced. We can observe that this symplectic  matrix is still singular, however we have showed that  there are not more constraints, therefore this theory is a  gauge theory. In order to obtain a symplectic tensor, we need to  fixing  the gauge, we will use the following temporal gauge
\begin{eqnarray}
A_{0}^{\ i}&=&0,  \\ \nonumber 
A_{00i}&=&0, \\ \nonumber
B_{0a}^{\ \ 0i}&=&0,  \\ \nonumber 
B_{0ai}&=&0 ,\label{gauge4}
\end{eqnarray}
this means  that
$\dot{T}^{i}=0, \dot{\Lambda}_i=0, \dot{\varsigma}_{a}^{\ i}=0$ and $\dot{\chi}_{ai}=0$. In this manner, we introduce more Lagrange multipliers enforcing the gauge fixing as constraints. The Lagrange multipliers introduced are
$\beta_i,\alpha^{i},\rho_{i}^{\ a},\sigma_{a}^{\ i}$ thus, the symplectic Lagrangian reads 
\begin{eqnarray}
\mathcal{L}^{(2)}&=&\Xi\eta^{abc}B_{ab}^{\ \ 0i}\dot{A}_{c0i}+\Xi\eta^{abc}B_{abi}\dot{\Upsilon}_{c}^{\ i}-\dot{T}^{i}\left[\Omega^{(0)}_{i}-\beta_i\right]-\dot{\Lambda}_i\left[\Omega^{(0) \ 00i}-\alpha^{i}\right]\nonumber \\
&&-\dot{\varsigma}_{a}^{\ i}\left[\frac{\Omega^{(0) \ 0a}_{\ \ \ \ \ \ 0i}}{2}-\rho_{i}^{\ a}\right]-\dot{\chi}_{ai}\left[\frac{\Omega^{(0) \ 0ai}}{2}-\sigma^{ai}\right]. \label{lagrangianosimplectico3-2}
\end{eqnarray}
From the symplectic Lagrangian (\ref{lagrangianosimplectico3-2}) we identify the following symplectic variables
\begin{equation}
\xi^{(0)}=(A_{a0i},B_{ab}^{\ \ 0i},\Upsilon_{a}^{\ i},B_{abi},A_{0}^{\ i},A_{00i},B_{a0}^{\ \ 0i},B_{0ai},T^{i},\Lambda_i,\varsigma_{a}^{\ i},\chi_{ai},\beta_i,\alpha^i,\rho_{i}^{\ a},\sigma^{ai} ),
\end{equation}
and the 1-forms
\begin{eqnarray}
a^{(0)}&=&\left(\Xi\eta^{abc}B_{ab}^{\ \ 0i},0,\Xi\eta^{abc}B_{abi},0,-\left[\Omega^{(0)}_{\ \ \ i}-\beta_i\right],-\left[\Omega^{(0) \ 00i}-\alpha^i\right],\right.\\
&&-\left.\left[\frac{\Omega^{(0) \ 0a}_{\ \ \ \ \ \ 0i}}{2}-\rho_{i}^{\ a}\right],-\left[\frac{\Omega^{(0) \ 0ai}}{2}-\sigma^{ai}\right] \right).
\end{eqnarray}
Thus, the symplectic matrix is given by
\begin{eqnarray}
f^{(2)}_{ij}=\left( \begin{array}{c c c c c}
0 & \Xi \eta^{abc}\delta^{i}_{k} & 0 & 0 & \Xi\eta^{abc}\epsilon_{j \ i}^{\ k}B_{ab}^{\ \ 0j} \\ 
 \Xi \eta^{abc}\delta^{i}_{k} & 0 & 0 & 0 & \Xi\eta^{abc}\epsilon_{k \ i}^{\ j}A_{c0j} \\
 0 & 0 & 0 & -\Xi\eta^{abc}\delta^{i}_{k} & -\Xi\eta^{abc}\epsilon^{j}_{\ ik}B_{abj} \\
 0 & 0 & \Xi \eta^{abc}\delta^{i}_{k} & 0 & -\Xi\eta^{abc}D^{i}_{kc} \\
 -\Xi \eta^{abc}\epsilon_{j \ i}^{\ k}B_{ab}^{\ \ 0j} & -\Xi \eta^{abc}\epsilon^{\ j}_{k \ i}A_{c0j} & \Xi \eta^{abc}\epsilon^{j}_{\ ik}B_{abj} & \Xi\eta^{abc}D^{i}
 _{kc} & 0\\
 -\Xi\eta^{abc}\epsilon^{ikj}B_{abj} & \Xi\eta^{abc}d^{i}_{kc} & -\Xi\eta^{abc}\epsilon^{i}_{\ jk}B_{ab}^{\ \ 0j} & -\Xi\eta^{abc}\epsilon^{ijk}A_{c0j} & 0\\
 \Xi\eta^{abc}d^{i}_{kc} & -\frac{\Xi}{2}\eta^{abc}\delta^{i}_{k} & \Xi\eta^{abc}\epsilon_{i \ k}^{\ j}A_{b0j} & 0 & 0\\
-\Xi\eta^{abc}\epsilon^{ijk}A_{b0j} & 0 & \Xi\eta^{abc}d^{i}_{kb} & -\Xi\eta^{abc}\delta^{i}_{k} & 0\\
0 & 0 & 0 & 0 & \delta^{i}_{j}\\
0 & 0 & 0 & 0 & 0\\
0 & 0 & 0 & 0 & 0\\
0 & 0 & 0 & 0 & 0\\
\end{array} \right. \nonumber 
\end{eqnarray}
\begin{eqnarray}
\left.\begin{array}{c c c c c c c}
\Xi\eta^{abc}\epsilon^{ikj}B_{abj} & -\Xi\eta^{abc}d^{i}_{kc} & \Xi\eta^{abc}\epsilon^{ijk}A_{b0j} & 0 & 0 & 0 & 0\\
-\Xi\eta^{abc}d^{i}_{kc} & \frac{\Xi}{2}\eta^{abc}\delta^{i}_{k} & 0 & 0 & 0 & 0 & 0\\
\Xi\eta^{abc}\epsilon^{i}_{\ jk}B_{ab}^{\ \ 0j} & -\Xi\eta^{abc}\epsilon_{i}^{\ jk}A_{b0j} & -\Xi\eta^{abc}d^{i}_{kb} & 0 & 0 & 0 & 0\\
\Xi\eta^{abc}\epsilon^{ijk}A_{c0j} & 0 & \frac{\Xi}{2}\eta^{abc}\delta^{i}_{k} & 0 & 0 & 0 & 0\\
0 & 0 & 0 & -\delta^{i}_{j} & 0 & 0 & 0\\
0 & 0 & 0 & 0 & -\delta^{i}_{j} & 0 & 0\\
0 & 0 & 0 & 0 & 0 & -\delta^{a}_{b}\delta^{i}_{j} & 0\\
0 & 0 & 0 & 0 & 0 & 0 & -\delta^{a}_{b}\delta^{i}_{j}\\
0 & 0 & 0 & 0 & 0 & 0 & 0\\
\delta^{i}_{j} & 0 & 0 & 0 & 0 & 0 & 0\\
0 & \delta^{a}_{b}\delta^{i}_{j} & 0 & 0 & 0 & 0 & 0\\
0 & 0 & \delta^{a}_{b}\delta^{i}_{j} & 0 & 0 & 0 & 0\\
\end{array} \right)\delta^3(x-y).\label{matrizsimplectica4}
\end{eqnarray}
We can observe that this matrix is not singular, after a long calculation, the inverse of $f_{ij}^{(2)}$ is given by
\begin{eqnarray}
f^{(2) \ -1}_{ij}=\left(\begin{array}{c c c c c c c c}
0 & \frac{1}{2\Xi}\eta_{abg}\delta^{k}_{l} & 0 & 0 & 0 & 0 & 0 & 0\\
-\frac{1}{2\Xi}\eta_{abg}\delta^{k}_{l} & 0 & 0 & 0 & 0 & 0 & 0 & 0\\
0 & 0 & 0 & \frac{1}{2\Xi}\eta_{abg}\delta^{k}_{l} & 0 & 0 & 0 & 0 \\
0 & 0 & -\frac{1}{2\Xi}\eta_{abg}\delta^{k}_{l} & 0 & 0 & 0 & 0 & 0\\
0 & 0 & 0 & 0 & 0 & 0 & 0 & 0\\
0 & 0 & 0 & 0 & 0 & 0 & 0 & 0\\
0 & 0 & 0 & 0 & 0 & 0 & 0 & 0\\
0 & 0 & 0 & 0 & 0 & 0 & 0 & 0\\
\epsilon_{l \ j}^{\ m}A_{g0m} & -\epsilon_{m}^{\ lj}B_{de}^{\ \ 0m} & -D_{gl}^{\ \ j} & \epsilon^{mj}_{\ \ l}B_{bgm} & -\delta^{k}_{i} & 0 & 0 & 0\\
-d_{g \ l}^{\ j} & -\epsilon^{jlm}B_{dem} & \epsilon^{jm}_{\ \ l}A_{g0m} & -\epsilon^{j}_{\ ml}B_{bg}^{\ \ 0m} & 0 & -\delta^{k}_{i} & 0 & 0\\
\frac{1}{2}\delta^{c}_{a}\delta^{b}_{g}\delta^{j}_{l} & \frac{1}{2}\delta^{de}_{ag}d_{e}^{\ lj} & 0 & -\frac{1}{2}\delta^{de}_{ag}\epsilon^{jm}_{\ \ l}A_{d0m} & 0 & 0 & -\delta^{c}_{a}\delta^{k}_{i} & 0\\
0 & -\frac{1}{2}\delta^{de}_{ag}\epsilon^{jml}A_{e0m} & -\frac{1}{2}\delta_{a}^{c}\delta^{j}_{l}\delta^{b}_{g} & \frac{1}{2}\delta^{de}_{ag}d_{d}^{\ lj} & 0 & 0 & 0 & -\delta_{a}^{c}\delta^{k}_{i}
\end{array} \right. \label{matrizinversa}.
\end{eqnarray}
\begin{eqnarray}
\left. \begin{array}{c c c c c c c}
0 & 0 & 0 & -\epsilon^{ij}_{\ \ l}A_{c0j} & d_{ci}^{\ \ l} & -\frac{1}{2}\delta_{a}^{f}\delta^{i}_{l} & 0 \\
0 & 0 & 0 & \epsilon_{jil}B_{ab}^{\ \ 0j} & \epsilon^{l \ j}_{\ i}B_{abj} & -\delta_{a}^{f}d_{b \ i}^{\ l} & \delta_{a}^{f}\epsilon^{lj}_{\ \ i} A_{b0j} \\
0 & 0 & 0 & D_{c}^{\ il} & -\epsilon^{lj}_{\ \ i} A_{c0j} & 0 & \frac{1}{2}\delta_{a}^{f}\delta_{il} \\
0 & 0 & 0 & -\epsilon^{jli} B_{abj} & \epsilon^{l}_{\ ji} B_{ab}^{\ \ 0j} & -\delta_{a}^{f}\epsilon^{lj}_{\ \ i} A_{b0j} & -\delta_{a}^{f}d_{bi}^{\ \ l} \\
0 & 0 & 0 & \delta^{i}_{j} & 0 & 0 & 0 \\
0 & 0 & 0 & 0 & \delta^{i}_{j} & 0 & 0 \\
0 & 0 & 0 & 0 & 0 & \delta^{a}_{b}\delta^{i}_{j} & 0 \\
0 & 0 & 0 & 0 & 0 & 0 & \delta^{a}_{b}\delta^{i}_{j} \\
0 & 0 & 0 & 0 & -G_{kl} & E^{k}_{\ l} & -H_{kl} \\
-\delta^{i}_{j} & 0 & 0 & C_{kl} & 0 & -I^{k}_{\ l} & -J^{k}_{\ l} \\
0 & -\delta^{a}_{b}\delta^{i}_{j} & 0 & E^{k}_{\ l} & -I^{k}_{\ l} & 0 & 0 \\
0 & 0 & -\delta^{a}_{b}\delta^{i}_{j} & H_{kl} & -J^{k}_{\ l} & 0 & 0 
\end{array}\right)\delta^3(x-y),  \label{matrizsimplectica5c}
\end{eqnarray}
where the following definitions have  been used 
\begin{eqnarray}
C_{kl}&=&\Xi \eta^{abc}\left( \epsilon_{jik}d_{c}^{\ il}B_{ab}^{\ \ 0j}-\epsilon^{jl}_{\ \ i}D^{i}_{\ kc}B_{abj} \right) , \nonumber \\
G_{kl}&=&\Xi \eta^{abc}\left( -d_{\ kb}^{i}D_{cil}+\epsilon_{\ \ k}^{ij}\epsilon_{i \ l}^{\ m}A_{b0j}A_{c0m}-\frac{1}{2}\epsilon_{\ lk}^{j}B_{abj} \right),\\
E_{kl}&=&\Xi \eta^{abc}\left( -\epsilon^{ij}_{\ \ k}A_{b0j}D_{c \ l}^{\ i}+\epsilon_{\ \ l}^{ij}A_{c0j}d_{ikc}-\frac{1}{2}\epsilon_{j}^{\ il}B_{ab}^{\ \ 0j} \right),\\
J_{kl}&=&\frac{\Xi}{2}\eta^{abc}\epsilon_{jkl}B_{ab}^{\ 0j},\\
I_{kl}&=&\Xi \eta^{abc}\left( d^{i}_{\ kb}d_{cil}-\epsilon^{\ j}_{l \ i}d^{i}_{\ kb}A_{c0j}+\frac{1}{2}\epsilon_{ljk}B_{ab}^{\ \ 0j} \right).
\end{eqnarray}
Therefore, from the symplectic tensor (\ref{matrizsimplectica5c}) we can identify the generalized FJ brackets by means of
\begin{equation}
\{\xi_{i}^{(2)}(x),\xi_{j}^{(2)}(y) \}_{FJ}=\left[f_{ij}^{(2)}(x,y) \right]^{-1},
\end{equation}
thus, the following generalized brackets arise
\begin{eqnarray}
\{B_{ab0i}(x),A_{d0l}(y) \}_{FJ}&=&\frac{1}{2\Xi}\eta_{abd}\eta_{il}\delta^{3}(x-y), \label{PFJ1} \\
\{B_{abi}(x),\Upsilon_{dl}(y) \}_{FJ}&=&- \frac{1}{2\Xi}\eta_{abd}\eta_{il}\delta^{3}(x-y), \label{PFJ2}
\end{eqnarray}
where we can observe that the FJ brackets depend  on the parameter $\Xi$. It is important to remark, that these brackets were not reported in \cite{19},   and these brackets will be useful in the quantization of theory.  As it was commented  previously, in the FJ formalism it is not necessary  the  classification of  the constraints in first class, second class, etc, such as in Dirac's method it is done; in FJ approach all constraints are treated at the same footing. In this manner,  the counting of physical degrees  of freedom is carried out  as follows  [DF=dynamical variables-independent constraints], thus,  for the theory under study, there are  18 canonical variables given by $(A_{c0i},\Upsilon_{c}^{\ i})$ and 18 independent constraints $(\Omega^{(0) \ i},\Omega^{(0) \ 00i},\Omega^{(0) \ 0ai},\Omega^{(0) \ 0a}_{\ \ \ \ \ \ 0i} )$, then the theory is devoid of physical degrees of freedom. This result  is expected because Pontryagin class is a topological theory. It is important to comment that all  these results are not reported in the literature.  \\
\textbf{A quantum state}\\
It is well-known from the quantum point of view, that  the  Dirac constraints   of the  Pontryagin class are solved by means the so-called  Chern-Simons state. Hence, in this section we will solve the quantum FJ constraints by using the  results reported above. We will observe that in spite of either Pontryagin or  Euler classes sharing the same classical equations of motion, their  respective quantum states will be different to each other. In order  to prove this claim   we need to rewrite the Chen- Simons action given by 
\begin{equation}
S[A]= \frac{\Xi}{2} \left[ \int  A^{IJ}\wedge A_{IJ} +  \frac{2}{3} A^{IK}\wedge A_{KL}\wedge A_{{I}}{^L}\right], \label{eq83}
\end{equation}
 in terms of the variables (\ref{variables11}), hence,   the action  takes the following form
\begin{equation}
S[A_a{^{0i}}, \Upsilon ^i_a]= \int \Bigg\{ \Xi \eta^{abc} \left[  A_a{^{0i}} \partial_b A_{c0i} + \Upsilon^k _a \partial_b \Upsilon_{ck} \right] -\Xi  \epsilon_{ijk} \eta^{abc} A_a{^{0i}}A_{c0}{^{j}} \Upsilon^k_b  +\frac{\Xi}{3} \epsilon_{ijk} \eta^{abc}\Upsilon^i_a \Upsilon ^j_b \Upsilon^k_c\Bigg\} dx^3.  \label{state}
\end{equation}
On the other hand, the generalized FJ brackets will be useful for the quantization. In fact, the dynamical variables will be promoved to operators and  the brackets  will be promoved to commutators. Hence, the generalized brackets are given by 
\begin{eqnarray}
\{B_{ab0i}(x),A_{d0l}(y) \}_{FJ}&=& \frac{1}{2\Xi}\eta_{abd}\eta_{il}\delta^{3}(x-y), \\
\{B_{abi}(x),\Upsilon_{dl}(y) \}_{FJ}&=&- \frac{1}{2\Xi}\eta_{abd}\eta_{il}\delta^{3}(x-y), 
\end{eqnarray}
its classical-quantum correspondence is given by 
\begin{eqnarray}
\{A_{d0i}(x),  \Xi \eta^{abg}\widehat{B}_{ab0j}(y) \}_{FJ}&=&-\eta_{ij}\delta^g_d\delta^{3}(x-y), \label{PFJ1} \\
\{\Upsilon_{di}(x), \Xi \eta^{abc} \widehat{B}_{abj}(y), \}_{FJ}&=& \eta_{ij} \delta^c_d\delta^{3}(x-y), 
\end{eqnarray}
therefore, we can identify the classical-quantum correspondence $\Xi \eta^{abg} \widehat{B}_{ab}{^{0j}} \rightarrow -  \mathbf{i} \frac{\delta}{\delta A_{g}{^{0j}} }$ and  $\Xi \eta^{abc} \widehat{B}_{abi}   \rightarrow   \mathbf{i} \frac{\delta}{\delta \Upsilon_{ci}} $. It is well-known, that   in theories with a  Hamiltonian described as  a linear combination of constraints as in our case, it is not possible  to use the Schrodinger equation for quantization,  because the action of the Hamiltonian on physical states is annihilation, in this manner, at quantum level we can not talk about the eigenstates of energy for the Hamiltonian \cite{19}. In canonical (symplectic)  quantization we have that the restriction of our physical states  is archived by demanding that
\begin{eqnarray}
  \Big\{\mathbf{i}  \frac{\delta}{\delta A_a{^{0i}}} &-& \Xi\eta^{abc} \left[ \partial_b A_{c0i} \partial_c A_{b0i}+ \epsilon_{i}{^{jk}} A_{b0j}\Upsilon_{ck} - \epsilon_{ijk}  A_{c0}{^{j}}\Upsilon_b^k  \right] \Big \}\Psi_P(A_a{^{0i}}, \Upsilon_a^i)=0, \nonumber\\
    \Big\{ \mathbf{i}  \frac{\delta}{\delta  \Upsilon_a^i} &-& \Xi \eta^{abc} \left[ \partial_b \Upsilon_c^i - \partial_c \Upsilon_b^i +\epsilon^{i}{_{jk}} \Upsilon^j_b \Upsilon^k_c-\epsilon^{ijk} A_{b0j}A_{c0k} \right]   \Big\}\Psi_P(A_a{^{0i}}, \Upsilon_a^i)=0, \nonumber \\
     \Big\{  \mathbf{i}  \partial_a  \frac{\delta}{\delta  \Upsilon_a^i} &+&   \mathbf{i}\epsilon^{j}{_{ik}}  \Upsilon_a^k  \frac{\delta}{\delta  \Upsilon_a^j} -  \mathbf{i} \epsilon_{jki} A_{c0}{^{k}}  \frac{\delta}{\delta  \Upsilon_a^j}\Big\} \Psi_P(A_a{^{0i}}, \Upsilon_a^i)=0, \nonumber \\
      \Big\{  \mathbf{i}  \partial_a  \frac{\delta}{\delta A_a{^{0i}}} &-&   \mathbf{i}\epsilon_ {i}{^{j}}{_{k}}  \Upsilon_a^k  \frac{\delta}{\delta A_a{^{0j}}} -  \mathbf{i} \epsilon_{i}{^j}{_k} A_{c0j}   \frac{\delta}{\delta  \Upsilon_a^j} \Big\} \Psi_P(A_a{^{0i}}, \Upsilon_a^i)=0, 
\end{eqnarray} 
where the solution is given by 
\begin{equation}
\Psi_P(A_a{^{0i}}, \Upsilon_a^i)= e^{- \mathbf{i} \Xi S[A_a{^{0i}}, \Upsilon_a^i]}, \label{wave1}
\end{equation}
here $S[A_a{^{0i}}, \Upsilon_a^i]$ is the action given in (\ref{state}). We can observe that the constraints are solved exactly and the  Bianchi identities are not involved;  this is a difference between our results and those reported in \cite{19}. In this manner, by using the new variables, the Chern-Simons state is a quantum state of the Pontryagin class. \\
\section{Symplectic analysis for the Euler invariant}
The Euler invariant is described by the the following action expressed as a  $BF$-like  theory
\begin{equation}
S[A_{\mu}^{\ IJ},B_{\alpha \beta}^{KL}]=\Omega \int_{M}^{}\left[\ast F^{IJ} \wedge B_{IJ}-\frac{1}{2}\ast B^{IJ} \wedge B_{IJ} \right]\label{euler11},
\end{equation}
where $\ast=\epsilon_{\ \ \ IJ}^{KL}$ is the dual of $SO(3,1)$ and  $\Omega$ is a constant. Both actions (\ref{chern2})  and (\ref{euler11}) give rise the same equations of motion. \\
 By performing the 3+1 decomposition,   breaking down the Lorentz covariance and using the variables (\ref{variables11}) we obtain the following Lagrangian density
\begin{eqnarray}
\mathcal{L}&=&-\Omega \eta^{abc}\dot{\Upsilon}_{a}^{\ j}B_{0jbc}+\Omega \eta^{abc}\dot{A}_{a \ l}^{\ 0}B_{bc}^{\ \ l} \nonumber \\
&&-A_{0}^{\ i}\left[\partial_a \left(\Omega \eta^{abc}B_{bc0i} \right)+\Omega \eta^{abc} \epsilon^{j}_{\ in}\Upsilon_{a}^{\ n}B_{0jbc}-\Omega \eta^{abc}\epsilon_{kli}A_{a}^{\ l0}B_{bc}^{\ \ k} \right] \nonumber \\
&&-A_{00l}\left[\partial_a \left(\Omega \eta^{abc}B_{bc}^{\ \ l} \right)-\Omega \eta^{abc}\epsilon_{\ in}^{l} \Upsilon_{a}^{\ n}B_{bc}^{\ \ i}+\Omega \eta^{abc}\epsilon^{jl}_{\ \ k}A_{a}^{\ 0k}B_{bc0j} \right] \nonumber \\
&&+\Omega \eta^{abc}B_{0c}^{\ \ 0j}\left[\partial_a \Upsilon_{bj}-\partial_b \Upsilon_{aj}-\epsilon_{jkl}A_{a \ 0}^{\ k}A_{b}^{\ 0l}+\epsilon_{jin}\Upsilon_{a}^{\ i}\Upsilon_{b}^{\ n} \right]\nonumber \\
&&+\Omega \eta^{abc}B_{0ci}\left[\partial_a A_{b}^{\ 0i}-\partial_b A_{a}^{\ 0i}-\epsilon^{mi}_{\ \ n}A_{a \ m}^{\ 0}\Upsilon_{b}^{\ n}+\epsilon^{i}_{\ mn}A_{b}^{\ m0}\Upsilon_{a}^{\ n} \right] \nonumber \\
&&+\Omega \eta^{abc}\left[B_{ab}^{\ \ i}B_{0c0i}+B_{0a}^{\ \ i}B_{bc0i} \right]\label{euler3}, 
\end{eqnarray}
where  $\epsilon^{0ijk} \equiv \epsilon^{ijk}$ and $i, j, k=1,2,3$ are  raised and  lowered with the Euclidean metric $\eta_{ij}=(1,1,1)$. We can see that either Euler or  Pontryagin theories share the  same configuration variables, however, the canonical partner's of the dynamical variables have changed, this fact will be reflected in the generalized FJ brackets.    In this respect, we will compare the results obtained from  both actions  (\ref{chern2}) and (\ref{euler11}), now   they are   at the same level written in terms of real variables, this is an different scenario to that  reported in \cite{19} where  it was considered only the auto-self-dual case.\\
In this manner, from (\ref{euler3})  we can identify the following symplectic Lagrangian given by
\begin{equation}
\mathcal{L}^{(0)}=\Omega\eta^{abc}B_{ab}^{\ \ 0i}\dot{\Upsilon}_{ci}-\Omega\eta^{abc}B_{abi}\dot{A}_{c0}^{\ \ i}-\mathcal{V}^{(0)}, \label{lagrangianosimplecticoE}
\end{equation}
where $\mathcal{V}^{(0)}$ is the symplectic potential 
\begin{eqnarray}
\mathcal{V}^{(0)}&=&A_{00i}\left[\partial_c(\Omega \eta^{abc}B_{abi} )+\Omega\eta^{abc}\epsilon^{j}_{\ ik}B_{abj}\Upsilon_{c}^{\ k}-\Omega\eta^{abc}\epsilon_{jki}B_{ab}^{\ \ 0j}A_{c0}^{\ \ k}\right]\nonumber \\
&&-A_{0}^{\ i}\left[\partial_c(\Omega \eta^{abc}B_{ab}^{\ \ 0i} )-\Omega \eta^{abc}\epsilon^{i}_{\ jk}B_{ab}^{\ \ 0j}\Upsilon_{c}^{\ k}-\Omega \eta^{abc} \epsilon^{ijk}B_{abk}A_{c0j} \right]
 \nonumber \\
&&+\Omega \eta^{abc}B_{0ai}\left[\partial_b A_{c0}^{\ \ i}-\partial_c A_{b0}^{\ \ i}+\epsilon^{ijk}A_{b0j}\Upsilon_{ck}-\epsilon_{\ jk}^{i}A_{c0}^{\ \ j}\Upsilon_{b}^{\ k}+B_{bc}^{\ \ 0i} \right] \nonumber \\
&&-\Omega \eta^{abc}B_{0a}^{\ \ 0i}\left[\partial_b \Upsilon_{ci}-\partial_c \Upsilon_{bi}+\epsilon_{ijk}\Upsilon_{b}^{\ j}\Upsilon_{c}^{\ k}-\epsilon^{\ jk}_{i}A_{b0j}A_{c0k}-B_{bcj} \right] .
 \end{eqnarray}
 Now, from the symplectic Lagrangian (\ref{lagrangianosimplecticoE}) we identify the following symplectic variables
 \begin{equation}
\xi^{(0)}=(A_{a0}^{\ \ i},B_{ab}^{\ \ 0i},\Upsilon_{ai},B_{abi},A_{0}^{\ i},A_{00i},B_{a0}^{\ \ 0i},B_{0ai}),
\end{equation} 
and the 1-form
\begin{equation}
a^{(0)}=(-\Omega\eta^{abc}B_{abi},0,\Omega\eta^{abc}B_{ab}^{\ \ 0i},0,0,0,0,0 ).
\end{equation} 
Thus, the symplectic matrix for the Euler class takes the  form
\begin{eqnarray}
f^{(0)}_{ij}=\left( \begin{array}{l l l l l l l l}
0 & 0 & 0 & \Omega \eta^{dec}\delta^{l}_{i} & 0 & 0 & 0 & 0\\ 
 0 & 0 & \Omega \eta^{dec}\delta^{l}_{i} & 0 & 0 & 0 & 0 & 0\\
 0 & -\Omega \eta^{dec}\delta^{l}_{i} & 0 & 0 & 0 & 0 & 0 & 0\\
 -\Omega \eta^{dec}\delta^{l}_{i} & 0 & 0 & 0 & 0 & 0 & 0 & 0\\
 0 & 0 & 0 & 0 & 0 & 0 & 0 & 0\\
 0 & 0 & 0 & 0 & 0 & 0 & 0 & 0\\
 0 & 0 & 0 & 0 & 0 & 0 & 0 & 0\\
 0 & 0 & 0 & 0 & 0 & 0 & 0 & 0
\end{array} \right)\delta^3(x-y). \label{matrizf0}
\end{eqnarray}
We observe that $f_{ij}^{(0)}$ is singular as expected because there are constraints. The zero-modes of $f_{ij}^{(0)}$ are given by the following 4 vectors
\begin{eqnarray}
&&\nu^{(0)}_1=(0,0,0,0,V^{A_{0}^{\ i}},0,0,0),\\
&&\nu^{(0)}_2=(0,0,0,0,0,V^{A_{a0i}},0,0),\\
&&\nu^{(0)}_3=(0,0,0,0,0,0,V^{B_{a0}^{\ \ 0i}},0),\\
&&\nu^{(0)}_4=(0,0,0,0,0,0,0,V^{B_{a0i}}),
\end{eqnarray}
where $V^{A_{0}^{\ i}}$, $V^{A_{a0i}}$, $V^{B_{a0}^{\ \ 0i}}$ and $V^{B_{a0i}}$ are arbitrary functions. Hence, by using these modes we find the following FJ constraints
\begin{eqnarray}
\Omega^{(0)}_{i}&=&\int_{}^{}\mathbf{d}^{3}x\nu^{(0) \ i}_1\frac{\delta}{\delta \xi^{(0) \ i}}\int_{}^{}\mathbf{d}^{3}y \mathcal{V}^{(0)}(\xi)=\int_{}^{}\mathbf{d}^{3}x V^{A_{0}^{\ i}}\frac{\delta}{\delta A_{0}^{\ i}}\int_{}^{}\mathbf{d}^{3}y \mathcal{V}^{(0)}(\xi)\nonumber \\
&=&\partial_c(\Omega \eta^{abc}B_{ab \ i}^{\ \ 0} )-\Omega \eta^{abc}\epsilon_{ijk}B_{ab}^{\ \ 0j}\Upsilon_{c}^{\ k}-\Omega \eta^{abc} \epsilon^{\ jk}_{i}B_{abk}A_{c0j}, \label{restriccion1E}\\
\Omega^{(0) \ 00i}&=&\int_{}^{}\mathbf{d}^{3}x\nu^{(0) \ i}_2\frac{\delta}{\delta \xi^{(0) \ i}}\int_{}^{}\mathbf{d}^{3}y \mathcal{V}^{(0)}(\xi)=\int_{}^{}\mathbf{d}^{3}xV^{A_{a0i}}\frac{\delta}{\delta A_{a0i}}\int_{}^{}\mathbf{d}^{3}y \mathcal{V}^{(0)}(\xi) \nonumber \\
&=&-\left[\partial_c(\Omega \eta^{abc}B_{abi} )+\Omega\eta^{abc}\epsilon^{j}_{\ ik}B_{abj}\Upsilon_{c}^{\ k}-\Omega\eta^{abc}\epsilon_{jki}B_{ab}^{\ \ 0j}A_{c0}^{\ \ k}\right],\label{restriccion2E}\\
\Omega^{(0) \ a0}_{\ \ \ \ \ \ i}&=&\int_{}^{}\mathbf{d}^{3}x\nu^{(0) \ i}_3\frac{\delta}{\delta \xi^{(0) \ i}}\int_{}^{}\mathbf{d}^{3}y \mathcal{V}^{(0)}(\xi)=\int_{}^{}\mathbf{d}^{3}xV^{B_{a0}^{\ \ 0i}}\frac{\delta}{\delta B_{a0}^{\ \ 0i}}\int_{}^{}\mathbf{d}^{3}y \mathcal{V}^{(0)}(\xi) \nonumber \\
&=&\Omega\eta^{abc}\left[\partial_b \Upsilon_{ci}-\partial_c \Upsilon_{bi}+\epsilon_{ijk}\Upsilon_{b}^{\ j}\Upsilon_{c}^{\ k}-\epsilon^{\ jk}_{i}A_{b0j}A_{c0k}-B_{bci} \right],\label{restriccion3E}\\
\Omega^{(0) \ 0a}_{\ \ \ \ \ \ 0i}&=&\int_{}^{}\mathbf{d}^{3}x\nu^{(0) \ i}_4\frac{\delta}{\delta \xi^{(0) \ i}}\int_{}^{}\mathbf{d}^{3}y \mathcal{V}^{(0)}(\xi)=\int_{}^{}\mathbf{d}^{3}x V^{B_{a0i}}\frac{\delta}{\delta B_{a0i}}\int_{}^{}\mathbf{d}^{3}y \mathcal{V}^{(0)}(\xi)\nonumber \\
&=&-\Omega \eta^{abc}\left[\partial_b A_{c0i}-\partial_c A_{b0i}+\epsilon_{i}^{\ jk}A_{b0j}\Upsilon_{ck}-\epsilon_{ijk}A_{c0}^{\ \ j}\Upsilon_{b}^{\ k}+B_{bc}^{\ \ 0i}\right] \label{restriccion4E},
\end{eqnarray}
and we observe that there are the following 6 reducibility condition between the constraints 
\begin{eqnarray}
\partial_a \Omega^{(0) \ a0}_{\ \ \ \ \ \ i} &=&\epsilon_{i}^{\ jk}\Upsilon_{ak}\Omega^{(0) \ a0}_{\ \ \ \ \ \ j}-\epsilon_{i}^{\ jk}A_{a0k}\Omega^{(0) \ 0a}_{\ \ \ \ \ \ 0j}-\frac{1}{2}\Omega^{(0) \ 00}{_{i}},  \\ \nonumber
\partial_a \Omega^{(0) \ 0a}_{\ \ \ \ \ \ 0i}&=&\epsilon_{i}^{\ jk}A_{a0k}\Omega^{(0) \ a0}_{\ \ \ \ \ \ j}+\epsilon_{i}^{\ jk}\Upsilon_{ak}\Omega^{(0) \ 0a}_{\ \ \ \ \ \ 0j}+\frac{1}{2}\Omega^{(0)}_{i}.
\end{eqnarray}
We can observe that either  Pontryagin or  Euler class share the same FJ constraints, however this fact does not guarantee  that these actions will be equivalent at  the quantum level, as we will see this point below.  Now, we shall observe if there are more constraints. For this aim, we use the expression   (\ref{terciarias}), where the symplectic matrix $\bar{f}_{ij}$ for the Euler theory  is given by
\begin{eqnarray}
\bar{f}_{ij}=\left(\begin{array}{c c c}
0 & 0 & 0 \\ 
0 & 0 & \Omega \eta^{abc}\delta^{j}_{i} \\
 0 & -\Omega \eta^{abc}\delta^{j}_{i} & 0 \\
 -\Omega \eta^{abc}\delta^{j}_{i} & 0 & 0 \\
 0 & 0 & 0 \\
 0 & 0 & 0 \\
 0 & 0 & 0 \\
 0 & 0 & 0 \\
 \Omega \eta^{abc}\epsilon_{ik}^{\ \ j}B_{abj} & \Omega \eta^{abc}(\delta_{ik}\partial_c-\epsilon_{ikj}\Upsilon_{c}^{\ j} )  & -\Omega \eta^{abc}\epsilon_{ijk}B_{ab}^{\ \ 0j} \\
 -\Omega\eta^{abc}\epsilon_{jki}B_{ab}^{\ \ 0j} & -\Omega\eta^{abc}\epsilon_{kji}A_{c0}^{\ \ j}  & \Omega\eta^{abc}\epsilon^{i \ j}_{\ k}B_{abj}\\
 -2\Omega\eta^{abc}\epsilon_{i}^{\ jk}A_{b0j}  & 0 & 2\Omega\eta^{abc}\left(\delta^{k}_{i}\partial_{b}+\epsilon_{ijk}\Upsilon_{b}^{j} \right)\\
2\Omega\eta^{abc}\left(\delta^{i}_{k}\partial_b-\epsilon^{ik}_{\ \ j}\Upsilon_{b}^{\ j} \right) & \Omega \eta^{abc}\delta^{i}_{k} & 2\Omega\eta^{abc}\epsilon^{ijk}A_{b0j} 
\end{array} \right.\nonumber 
\end{eqnarray}
\begin{eqnarray}
\left.\begin{array}{c c c c c}
\Omega \eta^{abc}\delta^{j}_{i} & 0 & 0 & 0 & 0\\
0 & 0 & 0 & 0 & 0\\
0 & 0 & 0 & 0 & 0\\
0 & 0 & 0 & 0 & 0\\
-\Omega\eta^{abc}\epsilon^{\ jk}_{i}A_{c0j} & 0 & 0 & 0 & 0\\
\Omega\eta^{abc}\left(\delta^{ik}\partial_b+\epsilon^{ki}_{\ \ j}\Upsilon_{c}^{\ j} \right) & 0 & 0 & 0 & 0\\
-\Omega \eta^{abc}\delta^{k}_{i} & 0 & 0 & 0 & 0\\
0 & 0 & 0 & 0 & 0\\
\end{array}\right) \delta^3(x-y).
\end{eqnarray}
The matrix $\bar{f}_{ij}$   has  the following null vectors
\begin{eqnarray}
&&\vec{\mathbb{V}}_1=(\epsilon_{ijl}A_{c0}^{\ \ j}V^{l}\ ,\epsilon_{jil}B_{ab}^{\ \ 0j}V^{l}\ ,\partial_c V_i-\epsilon_{ijl}\Upsilon_{c}^{\ l}V^{j}\ ,\epsilon^{j}_{\ li}B_{abj}V^{l},0,0,0,0,V^{i},0,0,0 ),\nonumber  \\
&&\vec{\mathbb{V}}_2= (-\left[\partial_c V_i+\epsilon^{l}_{\ ij}\Upsilon^{j}_{\ c}V_l\right] \ ,-\epsilon^{l \ j}_{i}B_{abj}V_l \ ,-\epsilon_{i}^{\ lj}A_{c0j}V_{l} \ ,-\epsilon_{lji}B_{ab}^{\ \ 0j}V^{l},0,0,0,0,0,V^{i},0,0), \nonumber \\
&&\vec{\mathbb{V}}_3=(V_i\ ,-2(\partial_b V_i+\epsilon_{l}^{\ ji}\Upsilon_{bj}V^{l}) \ ,0,-2\epsilon_{l \ i}^{\ j}A_{b0j}V^{l},0,0,0,0,0,0,V^{i},0),\nonumber \\
&&\vec{\mathbb{V}}_4=(0,2\epsilon^{lj}_{\ \ i}A_{b0j}V_l \ ,V_i\ ,2(\partial_b V_i-\epsilon_{lji}\Upsilon^{j}_{\ b}V^{l}),0,0,0,0,0,0,0,V^{i} ), 
\end{eqnarray}
where $V^l$'s are arbitrary functions.
On the other hand, $Z_{k}(\xi)$ is given by
\begin{eqnarray}
Z_i(\xi)&=&\left(\begin{array}{c}
\frac{\delta \mathcal{V^{(0)}}}{\delta \xi^{(0)}j}\\
0\\
0\\
0
\end{array} \right)\\
&=&\left(\begin{array}{c}
\Omega\eta^{abc}\epsilon_{i}^{\ kj}A_{0}^{\ i}B_{abj}-\Omega\eta^{abc}\epsilon_{jki}A_{00i}B_{ab}^{\ \ 0j}-2\Omega\eta^{abc}\partial_b B_{0ak}\\
+2\Omega\eta^{abc}\epsilon_{i}^{\ kj}B_{0ai}\Upsilon_{cj}+2\Omega\eta^{abc}\epsilon^{\ jk}_{i}B_{0a}^{\ \ 0i}A_{b0j}\\
\\
-\Omega\eta^{abc}\epsilon_{jk}^{\ \ i}A_{00i}A_{c0}^{\ \ k}+\Omega\eta^{abc}\partial_cA_{0i}+\Omega\eta^{abc}\epsilon_{ijk}A_{0}^{\ i}\Upsilon_{c}^{\ k}\\
\\
\Omega\eta^{abc}\epsilon_{ijk}A_{0}^{\ i}B_{ab}^{\ \ 0j}+\Omega\eta^{abc}\epsilon^{ji}_{\ \ k}A_{00i}B_{abj}-2\Omega\eta^{abc}\epsilon_{i}^{\ jk}A_{b0j}B_{0ai}\\
+2\Omega\eta^{abc}\partial_bB_{0a}^{\ \ 0i}-2\Omega\epsilon^{jk}_{i}\Upsilon_{b}^{\ j}B_{0a}^{\ \ 0i}\\
\\
-\Omega\eta^{abc}\partial_cA_{00}^{\ \ i}+\Omega\eta^{abc}\epsilon^{ji}_{\ \ k}A_{00i}\Upsilon_{c}^{\ k}+\Omega\eta^{abc}\epsilon^{\ jk}_{i}A_{0}^{\ i}A_{c0j}\\
\\
\Omega^{(0)}_{i}\\
\\
\Omega^{(0) \ 00i}\\
\\
\Omega^{(0) \ 0a}_{\ \ \ \ \ \ 0i}\\
\\
\Omega^{(0) \ 0ai}\\
\\
0\\
\\
0\\
\\
0\\
\\
0
\end{array}\right).
\end{eqnarray}
Hence, the contraction $\vec{\mathbb{V}}_{i}^{\mu}Z_{\mu}=0$ gives identities because this contraction is a linear combination of constraints. Therefore, there are no more FJ constraints.\\
Furthermore, we will add the constraints   (\ref{restriccion1E}-\ref{restriccion4E}) to the symplectic Lagrangian using the following Lagrange multipliers, namely 
$A_{0}^{\ i}=\dot{T}^{i},A_{00i}=\dot{\Lambda}_i,B_{0a}^{\ \ 0i}=\frac{\dot{\varsigma}_{a}^{\ i}}{2},B_{0ai}=\frac{\dot{\chi}_{ai}}{2} $. Thus, the symplectic Lagrangian takes the form
\begin{eqnarray}
\mathcal{L}^{(1)}&=&\Omega\eta^{abc}B_{ab}^{\ \ 0i}\dot{\Upsilon}_{ci}-\Omega\eta^{abc}B_{abi}\dot{A}_{c0}^{\ \ i}-\dot{T}^{i}\Omega^{(0)}_{i}+\dot{\Lambda}_i\Omega^{(0) \ 00i}-\frac{\dot{\varsigma}_{a}^{\ i}}{2}\Omega^{(0) \ 0a}_{\ \ \ \ \ \ 0i}\nonumber \label{lagrangianosimplectico2E} \\
&&+\frac{\dot{\chi}_{ai}}{2}\Omega^{(0) \ 0ai}-\mathcal{V}^{(1)}, \label{lagrangiano2}
\end{eqnarray}
where
$\mathcal{V}^{(1)}=\mathcal{V}^{(0)}|_{\Omega^{(0)}_{i},\Omega^{(0) \ 00i},\Omega^{(0) \ 0a}_{\ \ \ \ \ \ 0i},\Omega^{(0) \ 0ai}=0}=0$, this result is expected because Euler class is diffeomorphism covariant just like GR.\\
From the symplectic Lagrangian (\ref{lagrangiano2}) we identify the following symplectic variables
\begin{equation}
\xi^{(1)}=(A_{a0}^{\ \ i},B_{ab}^{\ \ 0i},\Upsilon_{ai},B_{abi},A_{0}^{\ i},A_{00i},B_{a0}^{\ \ 0i},B_{0ai},T^{i},\Lambda_i,\varsigma_{a}^{\ i},\chi_{ai} ),
\end{equation}
and the 1-forms
\begin{equation}
a^{(1)}=\left(-\Omega \eta^{abc}B_{abi},0,\Omega\eta^{abc}B_{ab}^{\ \ 0i},0,-\Omega^{(0)}_{\ \ \ i},+\Omega^{(0) \ 00i},-\frac{\Omega^{(0) \ 0a}_{\ \ \ \ \ \ 0i}}{2},+\frac{\Omega^{(0) \ 0ai}}{2} \right).
\end{equation}
Hence, the symplectic matrix has the following form
\begin{eqnarray}
f^{(1)}_{ij}=\left(\begin{array}{c c c c}
0 & 0 & 0 & \Omega \eta^{abc}\delta^{k}_{i} \\ 
 0 & 0 & \Omega \eta^{abc}\delta^{k}_{i} & 0\\
 0 & -\Omega \eta^{abc}\delta^{k}_{i} & 0 & 0 \\
 -\Omega \eta^{abc}\delta^{k}_{i} & 0 & 0 & 0\\
\Omega \eta^{abc}\epsilon_{ik}^{\ \ j}B_{abj} & \Omega \eta^{abc}d_{ikc}  & -\Omega \eta^{abc}\epsilon_{ijk}B_{ab}^{\ \ 0j} & -\Omega \eta^{abc}\epsilon_{i}^{\ jk}A_{c0j} \\
 -\Omega\eta^{abc}\epsilon_{jki}B_{ab}^{\ \ 0j} & -\Omega\eta^{abc}\epsilon_{kji}A_{c0}^{\ \ j}  & \Omega\eta^{abc}\epsilon^{i \ j}_{\ k}B_{abj} & \Omega\eta^{abc}D^{ik}_{\ \ c}\\
-\Omega\eta^{abc}\epsilon_{i}^{\ jk}A_{b0j}  & 0 & \Omega\eta^{abc}d^{k}_{\ ib} & -\frac{\Omega}{2}\eta^{abc}\delta^{k}_{i} \\
\Omega\eta^{abc}d^{i}_{\ kb} & \frac{\Omega}{2} \eta^{abc}\delta^{i}_{k} & \Omega\eta^{abc}\epsilon^{ijk}A_{b0j}  & 0
\end{array} \right. \nonumber
\end{eqnarray}
\begin{eqnarray}
\left. \begin{array}{c c c c}
-\Omega\eta^{abc}\epsilon_{ik}^{\ \ j}B_{abj} & \Omega\eta^{abc}\epsilon_{jki}B_{ab}^{\ \ 0j} & \Omega\eta^{abc}\epsilon_{i}^{\ jk}A_{b0j} & -\Omega\eta^{abc}d^{i}_{\ kb}\\
-\Omega\eta^{abc}d_{ikc} & \Omega\eta^{abc}\epsilon_{kji}A_{c0}^{\ \ j} & 0 & -\frac{\Omega}{2}\eta^{abc}\delta^{i}_{k} \\
\Omega\eta^{abc}\epsilon_{iik}B_{ab}^{\ \ 0j} & -\Omega\eta^{abc}\epsilon^{ji}_{\ \ k}B_{abj} & -\Omega\eta^{abc}d^{k}_{\ ib} & -\Omega\eta^{abc}\epsilon^{ijk}A_{b0j} \\
\Omega\eta^{abc}\epsilon_{i}^{\ jk}A_{c0j} & -\Omega\eta^{abc}D^{ik}_{\ \ c} & \frac{\Omega}{2}\eta^{abc}\delta^{k}_{i} & 0 \\
0 & 0 & 0 & 0\\
0 & 0 & 0 & 0\\
0 & 0 & 0 & 0\\
0 & 0 & 0 & 0\\
\end{array}\right) \delta^3 (x-y), \label{matrizsimplectica2E}
\end{eqnarray}
where we have defined
$D^{i}_{kc}=\delta^{i}_{k}\partial_c+\epsilon_{k}^{\ ij}\Upsilon_{cj}$ and $d^{i}_{kc}=\delta^{i}_{k}\partial_c-\epsilon_{k}^{\ ij}\Upsilon_{cj}$. This  matrix is singular and we have proved that there are not more constraints,  thus,  this theory has a gauge symmetry.  
In order to obtain a symplectic tensor, we fixing  the   temporal gauge just like was done for the Pontryagin invariant (\ref{gauge4}). In this manner, we introduce more Lagrange multipliers enforcing the gauge fixing as constraints. The Lagrange multipliers  are given by 
$\beta_i,\alpha^{i},\rho_{i}^{\ a},\sigma_{a}^{\ i}$, thus, the symplectic Lagrangian takes the form
\begin{eqnarray}
\mathcal{L}^{(2)}&=&\Omega\eta^{abc}B_{ab}^{\ \ 0i}\dot{\Upsilon}_{ci}-\Omega\eta^{abc}B_{abi}\dot{A}_{c0}^{\ \ i}-\dot{T}^{i}\left[\Omega^{(0)}_{i}-\beta_i\right]+\dot{\Lambda}_i\left[\Omega^{(0) \ 00i}+\alpha^{i}\right]\nonumber \\
&&-\dot{\varsigma}_{a}^{\ i}\left[\frac{\Omega^{(0) \ 0a}_{\ \ \ \ \ \ 0i}}{2}-\rho_{i}^{\ a}\right]+\dot{\chi}_{ai}\left[\frac{\Omega^{(0) \ 0ai}}{2}+\sigma^{ai}\right]. \label{lagrangianosimplectico3}
\end{eqnarray}
From the symplectic Lagrangian, we identify the following new set of symplectic variables
\begin{equation}
\xi^{(2)}=(A_{a0}^{\ \ i},B_{ab}^{\ \ 0i},\Upsilon_{ai},B_{abi},A_{0}^{\ i},A_{00i},B_{a0}^{\ \ 0i},B_{0ai},T^{i},\Lambda_i,\varsigma_{a}^{\ i},\chi_{ai},\beta_i,\alpha^i,\rho_{i}^{\ a},\sigma^{ai} ),
\end{equation}
and the 1-forms
\begin{eqnarray}
a^{(2)}&=&\left(-\Omega\eta^{abc}B_{abi},0,\Omega\eta^{abc}B_{ab}^{\ \ 0i},0,-\left[\Omega^{(0)}_{\ \ \ i}-\beta_i\right],\left[\Omega^{(0) \ 00i}+\alpha^i\right],\right.\\
&&-\left.\left[\frac{\Omega^{(0) \ 0a}_{\ \ \ \ \ \ 0i}}{2}-\rho_{i}^{\ a}\right],\left[\frac{\Omega^{(0) \ 0ai}}{2}+\sigma^{ai}\right] \right).
\end{eqnarray}
Thus, the symplectic matrix is given by
\begin{eqnarray}
f^{(2)}_{ij}=\left(\begin{array}{c c c c c}
0 & 0 & 0 & \Omega \eta^{abc}\delta^{k}_{i} & -\Omega\eta^{abc}\epsilon_{ik}^{\ \ j}B_{abj} \\ 
 0 & 0 & \Omega \eta^{abc}\delta^{k}_{i} & 0 & -\Omega\eta^{abc}d_{ikc} \\
 0 & -\Omega \eta^{abc}\delta^{k}_{i} & 0 & 0 & \Omega\eta^{abc}\epsilon_{iik}B_{ab}^{\ \ 0j} \\
 -\Omega \eta^{abc}\delta^{k}_{i} & 0 & 0 & 0 & \Omega\eta^{abc}\epsilon_{i}^{\ jk}A_{c0j} \\
\Omega \eta^{abc}\epsilon_{ik}^{\ \ j}B_{abj} & \Omega \eta^{abc}d_{ikc}  & -\Omega \eta^{abc}\epsilon_{ijk}B_{ab}^{\ \ 0j} & -\Omega \eta^{abc}\epsilon_{i}^{\ jk}A_{c0j} & 0 \\
 -\Omega\eta^{abc}\epsilon_{jki}B_{ab}^{\ \ 0j} & -\Omega\eta^{abc}\epsilon_{kji}A_{c0}^{\ \ j}  & \Omega\eta^{abc}\epsilon^{i \ j}_{\ k}B_{abj} & \Omega\eta^{abc}D^{ik}_{\ \ c} & 0 \\
-\Omega\eta^{abc}\epsilon_{i}^{\ jk}A_{b0j}  & 0 & \Omega\eta^{abc}d^{k}_{\ ib} & -\frac{\Omega}{2}\eta^{abc}\delta^{k}_{i} & 0 \\
\Omega\eta^{abc}d^{i}_{\ kb} & \frac{\Omega}{2} \eta^{abc}\delta^{i}_{k} & \Omega\eta^{abc}\epsilon^{ijk}A_{b0j}  & 0 & 0 \\
0 & 0 & 0 & 0 & \delta^{i}_{j} \\
0 & 0 & 0 & 0 & 0 \\
0 & 0 & 0 & 0 & 0 \\
0 & 0 & 0 & 0 & 0
\end{array} \right. \nonumber
\end{eqnarray}
\begin{eqnarray}
\left. \begin{array}{c c c c c c c}
\Omega\eta^{abc}\epsilon_{jki}B_{ab}^{\ \ 0j} & \Omega\eta^{abc}\epsilon_{i}^{\ jk}A_{b0j} & -\Omega\eta^{abc}d^{i}_{\ kb} & 0 & 0 & 0 & 0 \\
\Omega\eta^{abc}\epsilon_{kji}A_{c0}^{\ \ j} & 0 & -\frac{\Omega}{2}\eta^{abc}\delta^{i}_{k} & 0 & 0 & 0 & 0 \\
-\Omega\eta^{abc}\epsilon^{ji}_{\ \ k}B_{abj} & -\Omega\eta^{abc}d^{k}_{\ ib} & -\Omega\eta^{abc}\epsilon^{ijk}A_{b0j} & 0 & 0 & 0 & 0 \\
-\Omega\eta^{abc}D^{ik}_{\ \ c} & \frac{\Omega}{2}\eta^{abc}\delta^{k}_{i} & 0 & 0 & 0 & 0 & 0 \\
0 & 0 & 0 & -\delta^{i}_{j} & 0 & 0 & 0 \\
0 & 0 & 0 & 0 & -\delta^{i}_{j} & 0 & 0 \\
0 & 0 & 0 & 0 & 0 & -\delta^{a}_{b}\delta^{i}_{j} & 0 \\
0 & 0 & 0 & 0 & 0 & 0 & -\delta^{a}_{b}\delta^{i}_{j} \\
0 & 0 & 0 & 0 & 0 & 0 & 0 \\
\delta^{i}_{j} & 0 & 0 & 0 & 0 & 0 & 0 \\
0 & \delta^{a}_{b}\delta^{i}_{j} & 0 & 0 & 0 & 0 & 0 \\
0 & 0 & \delta^{a}_{b}\delta^{i}_{j} & 0 & 0 & 0 & 0 
\end{array}\right)\delta^3(x-y). \label{matrizsimplectica4E}
\end{eqnarray}
We can observe that this matrix is not singular;  after a long calculation, the inverse of $f_{ij}^{(2)}$ is given by
\begin{eqnarray}
f^{(2) \ -1}_{ij}=\left(\begin{array}{c c c c c}
0 & 0 & 0 & -\frac{1}{2\Omega} \eta_{abg}\delta^{l}_{k} & 0 \\ 
 0 & 0 & -\frac{1}{2\Omega} \eta_{abg}\delta^{l}_{k} & 0 & 0 \\
 0 & \frac{1}{2\Omega} \eta_{abg}\delta^{l}_{k} & 0 & 0 & 0 \\
 \frac{1}{2\Omega} \eta_{abg}\delta^{l}_{k} & 0 & 0 & 0 & 0 \\
0 & 0  & 0 & 0 & 0 \\
 0 & 0 & 0 & 0 & 0 \\
0 & 0 & 0 & 0 & 0 \\
0 & 0 & 0  & 0 & 0 \\
-\epsilon_{j}^{\ kl}A_{g0k} & -\epsilon^{j \ l}_{\ m}B_{bg}^{\ \ 0m} & -d^{jl}_{\ \ g} & -\epsilon^{jlm}B_{bgm} & -\delta^{i}_{j} \\
D^{jl}_{\ \ g} & \epsilon^{jlm}B_{bgm} & \epsilon^{lmj}A_{g0j} & \epsilon_{m}^{\ lj}B_{bg}^{\ \ 0j} & 0 \\
-\frac{1}{2}\delta^{j}_{l}\delta^{c}_{a}\delta^{b}_{a} & \frac{1}{2}\delta^{fe}_{ag}D_{f}^{\ jl} & 0 & \frac{1}{2}\delta^{fe}_{ag}\epsilon^{jmk}A_{f0m} & 0 \\
0 & \frac{1}{2}\delta^{fe}_{ag}\epsilon^{jlm}A_{f0m} & -\frac{1}{2}\delta^{c}_a \delta^{b}_{g}\delta^{jl} & -\frac{1}{2}\delta^{fe}_{ag}d^{jl}_{\ \ f} & 0
\end{array} \right. \nonumber
\end{eqnarray}
\begin{eqnarray}
\left. \begin{array}{c c c c c c c}
0 & 0 & 0 & \epsilon_{l}^{\ ji}A_{c0j} & -D_{l \ c}^{\ i} & \frac{1}{2}\delta_{a}^{f}\delta^{i}_{l} & 0 \\
0 & 0 & 0 & \epsilon_{lji}B_{ab}^{\ \ 0j} & -\epsilon^{jl}_{\ i}B_{abj} & -\delta_{a}^{f}D_{lib} & \delta_{a}^{f}\epsilon_{l \ i}^{\ j}A_{b0j} \\
0 & 0 & 0 & d_{lic} & -\epsilon_{i \ l}^{\ j}A_{c0j} & 0 & \frac{1}{2}\delta_{a}^{f}\delta_{il} \\
0 & 0 & 0 & \epsilon_{l}^{\ ij}B_{abj} & -\epsilon_{j \ l}^{\ i}B_{ab}^{\ \ 0j} & -\delta_{a}^{f}\epsilon_{l \ i}^{\ j}A_{b0j} & \delta_{a}^{f}d_{l \ b}^{\ i} \\
0 & 0 & 0 & \delta^{i}_{j} & 0 & 0 & 0 \\
0 & 0 & 0 & 0 & \delta^{i}_{j} & 0 & 0 \\
0 & 0 & 0 & 0 & 0 & \delta^{a}_{b}\delta^{i}_{j} & 0 \\
0 & 0 & 0 & 0 & 0 & 0 & \delta^{a}_{b}\delta^{i}_{j} \\
0 & 0 & 0 & 0 & C_{kl} & E^{k}_{\ l} & -I_{kl} \\
-\delta^{i}_{j} & 0 & 0 & C_{kl} & 0 & I^{k}_{\ l} & -E^{k}_{\ l} \\
0 & -\delta^{a}_{b}\delta^{i}_{j} & 0 & E^{k}_{\ l} & I^{k}_{\ l} & 0 & 0 \\
0 & 0 & -\delta^{a}_{b}\delta^{i}_{j} & G_{kl} & E^{k}_{\ l} & K_{kl}^{\ \ f} & 0 
\end{array}\right)\delta^3(x-y), \label{matrizsimplectica5E}
\end{eqnarray}
where we have defined
\begin{eqnarray}
C_{kl}&=&\Omega \eta^{abc}\left( \epsilon^{i \ j}_{k}B_{abj}d_{lic}+\epsilon_{l}^{\ ij}B_{abj}D^{i}_{\ kc} \right), \\
E_{kl}&=&\Omega \eta^{abc}\left( d_{lic}D_{kib}-\epsilon_{i}^{\ jk}\epsilon_{l}^{\ mi}A_{b0j}A_{c0m}-\frac{1}{2}\epsilon_{lk}^{j}B_{abj} \right),\\
G_{kl}&=&\Omega \eta^{abc}\left( \epsilon^{ij}_{\ \ k}A_{b0j}d_{lic}+\epsilon_{li}^{\ j}A_{c0j}d^{i}_{\ kb}+\frac{1}{2}\epsilon_{ljk}B_{ab}^{\ \ 0j} \right),\\
I_{kl}&=&\frac{\Omega}{2}\eta^{abc}\epsilon_{jkl}B_{ab}^{\ 0j},\\
K_{kl}^{\ \ f}&=&\Omega \eta^{fbc}\left( d_{lkb}-D_{lkb} \right).
\end{eqnarray}
Therefore, from the symplectic tensor (\ref{matrizsimplectica5E}) we can identify the following FJ brackets 
\begin{eqnarray}
\{B_{ab0i}(x),\Upsilon_{dl}(y) \}_{FJ}&=&\frac{1}{2\Omega}\eta_{abd}\eta_{il}\delta^{3}(x-y), \label{PFJE1} \\ 
\{B_{abi}(x),A_{d0l}(y) \}_{FJ}&=&\frac{1}{2\Omega}\eta_{abd}\eta_{il}\delta^{3}(x-y), \label{PFJE2}
\end{eqnarray}
which have been not reported in the literature. It is important to observe that in spite of we have used in  both theories  the same configuration variables and  the same gauge fixing, the generalized brackets  are different to each other. As it is showed  below  this fact will be important in the quantization .\\  
Now, the counting of physical degrees of freedom is performed in the following way  [DF=dynamical variables-independent constraints], thus, there are  18 canonical variables given by $(A_{c0i},\Upsilon_{c}^{\i})$ and 18 independent constraints $(\Omega^{(0) \ i},\Omega^{(0) \ 00i},\Omega^{(0) \ 0ai},\Omega^{(0) \ 0a}_{\ \ \ \ \ \ 0i} )$. In this manner, the theory  lacks of physical  degrees of freedom.  \\
$\textbf{A quantum state}$\\
We have seen in previous sections that   the   FJ constraints for the  Pontryagin theory  are exactly solved by means the so-called  Chern-Simons state. In this section we will solve the quantum FJ constraints  for the   Euler class.  First we observe that the state given in (\ref{wave1}) does not solve the Euler constraints, in this manner we need to find a new state.  We propose the following   Chen-Simons action 
\begin{equation}
S[A]= \frac{\Omega}{2} \left[ \int \epsilon^{IJKL} A_{IJ}\wedge A_{KL} +  \frac{2}{3} \epsilon^{IJKL} A_{I}{^{E}} \wedge A_{EJ}\wedge A_{KL}\right], \label{eq83}
\end{equation}
and we write it   in terms of the variables (\ref{variables11}), then  it takes the following form
\begin{equation}
S[A_a{^{0i}}, \Upsilon ^i_a]= \int \Bigg\{ \Omega \eta^{abc} \left[  -A_a{^{0i}} \partial_b \Upsilon_{c}{^{i}}   + A_a{^{0i}} \partial_c \Upsilon_{b}{^{i}}  \right] -\Omega  \eta^{abc}\epsilon_{ijk} A_{a0}{^{i}}   \Upsilon^j_b \Upsilon^k_c + \frac{\Omega}{3} \eta^{abc} \epsilon^{ijk}   A_{a0i}A_{b0j} A_{c0k}\Bigg\} dx^3.  \label{state2}
\end{equation}
On the other hand, the generalized FJ brackets for the Euler invariant will be useful for the quantization. In fact, the dynamical variables will be promoved to operators and the brackets  will be promoved to commutators. Hence, the generalized brakets are given by 
\begin{eqnarray}
\{B_{ab0i}(x),\Upsilon_{dl}(y) \}_{FJ}&=&\frac{1}{2\Omega}\eta_{abd}\eta_{il}\delta^{3}(x-y), \label{PFJE1} \\ 
\{B_{abi}(x),A_{d0l}(y) \}_{FJ}&=&\frac{1}{2\Omega}\eta_{abd}\eta_{il}\delta^{3}(x-y), \label{PFJE2}
\end{eqnarray}
its classical-quantum correspondence is given by 
\begin{eqnarray}
\{ \Upsilon_{di}(x), \Omega \eta^{abc} \widehat{B}_{bc0j}(y) \}_{FJ}&=&- \eta_{ij} \delta^a{_{d}}\delta^{3}(x-y), \label{PFJE1} \\ 
\{  A_{d0i}(x), \Omega \eta^{abc}  \widehat{B}_{bcj}(y) \}_{FJ}&=& -\eta_{ij} \delta^a{_{d}}\delta^{3}(x-y), \label{PFJE2}
\end{eqnarray}
hence, we can identify the classical-quantum correspondence $\Omega \eta^{abc} \widehat{B}_{bc0i} \rightarrow -\mathbf{i} \frac{\delta}{\delta \Upsilon_{a}^i}$ and  $\Omega \eta^{abc} \widehat{B}_{bci}   \rightarrow   - \mathbf{i} \frac{\delta}{\delta A_{a}{^{0i}} } $;  this election has been used because both  $A_a{^{0i}}$ and  $ \Upsilon_a^i $ are now  the dynamical variables. Moreover, just like in previously sections,  we will demand   that the restriction for the Euler   physical states, namely $\Psi_E(A_a{^{0i}}, \Upsilon_a^i)$,  will be  archived by 
\begin{eqnarray}
  \Big\{\mathbf{i}  \frac{\delta}{\delta A_a{^{0i}}} &-& \Omega \eta^{abc} \left[ \partial_b \Upsilon_c^i - \partial_c \Upsilon_b^i +\epsilon^{i}{_{jk}} \Upsilon^j_b \Upsilon^k_c-\epsilon^{ijk} A_{b0j}A_{c0k} \right]  \Big \}\Psi_E(A_a{^{0i}}, \Upsilon_a^i)=0, \nonumber\\
    \Big\{ \mathbf{i}  \frac{\delta}{\delta  \Upsilon_a^i} &-& \Omega \eta^{abc}  \left[ \partial_b A_{c0i} \partial_c A_{b0i}+ \epsilon_{i}{^{jk}} A_{b0j}\Upsilon_{ck} - \epsilon_{ijk}  A_{c0}{^{j}}\Upsilon_b^k  \right]   \Big\}\Psi_E(A_a{^{0i}}, \Upsilon_a^i)=0, \nonumber \\
     \Big\{ - \mathbf{i}  \partial_a  \frac{\delta}{\delta  \Upsilon_a^i} &+&   \mathbf{i} \Omega \epsilon^{i}{_{jk}}  \Upsilon_a^k  \frac{\delta}{\delta  \Upsilon_a^j} + \mathbf{i} \Omega \epsilon^{ijk} A_{a0j}  \frac{\delta}{\delta  A_{a}{^{0k}}}\Big\} \Psi_E(A_a{^{0i}}, \Upsilon_a^i)=0, \nonumber \\
      \Big\{-  \mathbf{i}  \partial_a  \frac{\delta}{\delta A_a{^{0i}}} &-&   \mathbf{i} \Omega \epsilon^{j}{_{ik}}  \Upsilon_a^k  \frac{\delta}{\delta A_a{^{0j}}} -  \mathbf{i} \Omega \epsilon_{ijk} A_{a0}{^{k}}   \frac{\delta}{\delta  \Upsilon_a^j} \Big\} \Psi_E(A_a{^{0i}}, \Upsilon_a^i)=0, \label{constraintseu}
\end{eqnarray} 
where the solution is given by 
\begin{equation}
\Psi_E(A_a{^{0i}}, \Upsilon_a^i)= e^{\mathbf{i} \Omega  S[A_a{^{0i}}, \Upsilon_a^i]}, \label{wave2}
\end{equation}
now  $S[A_a{^{0i}}, \Upsilon_a^i]$ is  given in (\ref{state2}). Again, the constraints (\ref{constraintseu}) are solved exactly by (\ref{wave2}), thus,  a  new quantum state is reported in this work.   In this manner, in spite of the Euler and  Pontryagin   sharing the same equations of motion, its  corresponding quantum states  are different. 
\section{Conclusions}
In this paper, a complete symplectic analysis for the Euler and Pontryaging invariants  has been performed. We carry out our analysis for both invariants  by using the same symplectic variables and the same gauge fixing,   we have observed that in spite of the Euler and  Pontryagin   invariants sharing the same FJ constraints, its  corresponding generalized FJ brackets are different. This fact, allowed us to observe that the solution to the quantum FJ constraints are not the same.  It is worth  to comment,  that we have found only mathematical solutions for the constraints;  in order to observe if these solutions are physical (we need to remember that we have worked with real variables) then it is necessary to construct a  measure for the quantization via mechanical path integral. In fact, there is an important connection between FJ quantization and path integral as that reported in \cite{23}. In this respect, the measure acquires a factor related with the determinant of the symplectic tensors  given in  (\ref{matrizsimplectica5c}) and (\ref{matrizsimplectica5E}), thus, in this paper we have all tools for exploring these subjects. Finally, we have seen that the FJ formalism demands to work with  less constraints than Dirac's formalism, this fact allowed us to construct the fundamental brackets  with  relative simplicity. Moreover, it is  possible to analice the addition of  the topological  invariants  to theories with  degrees of freedom    just like  bi-gravity models  \cite{24},  these problems are already in progress   and will be the subject of  forthcoming works \cite{25}. 


\section{References}
\begin{thebibliography}{}

\bibitem{1} C. Rovelli.  Quantum Gravity. Cambridge University Press, Cambridge (2004), T. Thiemann, Modern Canonical Quantum General Relativity. Cambridge University Press, Cambridge (2007).
\bibitem{2} E. Witten, J. Diff. Geom., 17 (4), 661-6692, (1982).
\bibitem{3} M. F. Atiyah, Publications Mathématiques de l'IHÉS, Volume 68,  175-186, (1988).
\bibitem{4}  E. Witten, Quantum field theory and the Jones polynomial, Commun. Math. Phys. 121, No. 3, 351-399,  (1989);  E. Witten, Topological quantum field theory, Comm. Math. Phys., 117,  353-386, (1988); E. Witten, Topological sigma models, Comm. Math. Phys., 118 (1988), 411-449. 
\bibitem{5} D.  J.  Rezende and   A. Perez, Phys. Rev. D 79: 064026, (2009).
\bibitem{6} H.  G.  Comp\'ean, O.  Obreg\'on, C. Ram{\'i}rez and  M. Sabido, Journal of Physics. Conference Series 24,  203-212, (2005).
\bibitem{7} A. Mardones, J. Zanelli, Class. Quantum Grav. 8 (1991) 1545.
 \bibitem{8}  T. Kimura, Prog. Theor. Phys. 42, 1191, (1969).
\bibitem{9}  R. Delbourgo, A. Salam, Phys. Lett. B 40  381, (1972).
\bibitem{10} T. Eguchi, P. Freund, Phys. Rev. Lett. 37  1251, (1976).
\bibitem{11}  L. Alvarez-Gaumé, E. Witten, Nucl. Phys. B 234,   269, (1984).
\bibitem{12}  O. Chandia, J. Zanelli, Phys. Rev. D 55, 7580, (1997).
\bibitem{13}  G.T. Horowitz, Commun. Math. Phys. 125, 417, (1989).
\bibitem{14}  G.T. Horowitz, M. Srednicki, Commun. Math. Phys. 130,  83, (1990).
\bibitem{15} J. F. Plebanski, J. Math. Phys. 18,  2511, (1977).
\bibitem{16}  M. Celada, M. Montesinos, J.  Romero, Class. Quant. Grav. 33, No 11, 115014, (2016);   D. K. Wise, Class .Quant. Grav. 27, 155010, (2010); 
\bibitem{17}  J.A. Nieto, J. Socorro, Phys. Rev. D59:
 041501, (1999).
\bibitem{18} J. C. Baez, Lect. Notes Phys. 543, 25-94,  (2000); A. Perez,  Living Rev. Relativ. (2013) 16: 3. https://doi.org/10.12942/lrr-2013-3. 
\bibitem{19} A. Escalante,  Phys.  Lett.  B 676,   105-111, (2009); I. Oda,  arXiv:hep-th/0311149.
\bibitem{20} O. Chand{\'i}a and J. Zanelli, AIP Conference Proceedings 419, 251 (1998); doi: http://dx.doi.org/10.1063/1.54694
\bibitem{21} A. Escalante and L. Carbajal, Annals  Phys. 326,  323?339, (2011).
\bibitem{22} L.D. Faddeev, R. Jackiw, Phys. Rev. Lett. 60 (1988) 1692.;  E.M.C. Abreu, A.C.R. Mendes, C. Neves, W. Oliveira, F.I. Takakura, L.M.V. Xavier, Modern Phys. Lett. A 23 (2008) 829;
E.M.C. Abreu, A.C.R. Mendes, C. Neves, W. Oliveira, F.I. Takakura, Internat. J. Modern Phys. A 22 (2007) 3605;
E.M.C. Abreu, C. Neves, W. Oliveira, Internat. J. Modern Phys. A 21 (2008) 5329;
C. Neves, W. Oliveira, D.C. Rodrigues, C. Wotzasek, Phys. Rev. D 69 (2004) 045016; J. Phys. A 3 (2004) 9303;
C. Neves, C. Wotzasek, Internat. J. Modern Phys. A 17 (2002) 4025;
C. Neves, W. Oliveira, Phys. Lett. A 321 (2004) 267;
J.A. Garcia, J.M. Pons, Internat. J. Modern Phys. A 12 (1997) 451;
E.M.C. Abreu, A.C.R. Mendes, C. Neves, W. Oliveira, R.C.N. Silva, C. Wotzasek, Phys. Lett. A 374 (2010) 3603?3607; A. Escalante, M. Z\'arate, Annals   Phys.  353 (2015) 163-178.
\bibitem{23} D. J. Toms, Phys. Rev. D, 92, 105026, (2015).
\bibitem{24} C. Deffayet, J. Mourad and G. Zahariade, JHEP, 03, 086, (2013); Tuan Q. Do. Phys. Rev. D 94, 044022 (2016). 
\bibitem{25} A. Escalante, work in progress. 
\end{thebibliography}
\end{document}